\documentclass[prd, preprint, longbibliography, 12pt]{revtex4-1}

\usepackage{amsmath}
\usepackage{amssymb}
\usepackage{setspace}
\usepackage{graphicx}
\usepackage{natbib}
\usepackage{float}
\usepackage{tensor}
\usepackage[utf8]{inputenc}
\usepackage{amsfonts}
\usepackage{fdsymbol}
\usepackage{braket}

\begin{document}
 
 %

\begin{center}
 { \large {\bf From Quantum Foundations,  to Spontaneous Quantum Gravity}}\\
{- {\it An overview of the new theory - }}


\vskip 0.3 in

{\large{\bf Tejinder P.  Singh}}

{\it Tata Institute of Fundamental Research,}
{\it Homi Bhabha Road, Mumbai 400005, India}\\
\bigskip
{\tt tpsingh@tifr.res.in}

\end{center}

\bigskip
\bigskip

\centerline{\bf ABSTRACT}
\noindent Spontaneous localisation is a falsifiable dynamical mechanism which modifies quantum mechanics,  and explains the absence of position superpositions in the macroscopic world. However, this is an ad hoc phenomenological proposal. Adler's theory of trace dynamics, working on a flat Minkowski space-time, derives quantum (field)  theory, and spontaneous localisation, as a thermodynamic approximation to an underlying noncommutative matrix dynamics. We describe how to incorporate gravity into trace dynamics, by using ideas from Connes' non-commutative geometry programme. This leads us to a new quantum theory of gravity, from which we can predict spontaneous localisation, and give an estimate of the Bekenstein-Hawking entropy of a Schwarzschild black hole.

\bigskip
\bigskip

\tableofcontents

\section{Introduction}
We have recently proposed  a new candidate quantum theory of gravity \cite{maithresh2019}, which we have named Spontaneous Quantum Gravity. The theory is built on the following principle: There ought to exist a reformulation of quantum (field) theory which does not depend on classical time \cite{Singh:2012}. Such a  time, as well as space-time, are properties of a  universe dominated by macroscopic material bodies. The space-time manifold as well as its pseudo-Riemannian geometry, in a classical universe, are determined by macroscopic matter fields. These large bodies are in turn a limiting case of a quantum description of matter fields. In the absence of such classical bodies, one cannot meaningfully talk of space-time geometry nor a space-time manifold. And yet there ought to be a way to describe quantum dynamics, say soon after the Big Bang, when nothing was classical. Hence, there ought to exist a reformulation of quantum theory which makes no reference to classical time. Such a reformulation is naturally also a quantum theory of gravity. As and when classical space-time and a universe dominated by macroscopic objects is recovered, this reformulation becomes equivalent to standard quantum field theory on a  background space-time. The development of such a reformulation is different from `quantization of the gravitational field' as we will see in more detail below. The latter procedure amounts to applying the rules of quantization to space-time geometry. Whereas our proposal is that since quantum rules are dependent on time, and hence in a sense dependent on their own limit, we must find a more precise formulation for them, which does not depend on the very limit of the theory.

We have developed such a reformulation of quantum theory [the afore-said spontaneous quantum  gravity] by building on Stephen Adler's theory of trace dynamics \cite{Adler:04}, and on Alain Connes' programme of non-commutative geometry \cite{Connes2000}. This has led us to a theory on the Planck scale, more general  than quantum theory, and to which relativistic quantum theory is a low-energy approximation. The mathematical structure of the theory is relatively straightforward to describe. Suppose one decides to do classical dynamics not with c-number valued matter degrees of freedom, but with [Grassmann valued] matrices. The Lagrangian of the theory then becomes a matrix polynomial. One takes the matrix trace of this polynomial  to construct the trace Lagrangian of the new matrix dynamics, assumed to operate on the Planck scale. Time integral of this trace Lagrangian [or four-volume integral in the case of a continuum matrix dynamics] yields the action function of the theory. The variation of this action with respect to the matrix valued matter degrees of freedom gives rise to a (Lorentz invariant)  Lagrangian dynamics, distinctly richer than ordinary classical dynamics. Space-time is still assumed to be Minkowski flat [as a simplifying approximation] even though the theory is assumed to operate at the Planck scale. This matrix dynamics is not quantum theory: the commutation relations amongst dynamical variables are time-dependent, and are not the commutation relations of quantum theory. The theory is invariant under global unitary transformations of the matrix-valued dynamical variables. The Hamiltonian of the theory, in general, is not self-adjoint. 

One next asks, what will this deterministic dynamics look like, if observed not on the Planck energy/time scale, but at much lower energies/time resolution? In other words, one is coarse-graining the evolution over time intervals much larger than Planck time, thereby smoothing out over the so-called space-time foam.  The emergent dynamics is quantum field theory, so long as the anti-self-adjoint component of the underlying Hamiltonian is negligible. Hence one says that quantum theory is an emergent phenomenon. If a large number of degrees of freedom get entangled [in the sense of quantum entanglement], the anti-self-adjoint component of the underlying Hamiltonian becomes significant, and the approximation necessary for emergence of quantum theory breaks down. Rapid spontaneous localisation results, leading to a breakdown of quantum superposition, and emergence of classical dynamics. 

This above, is the essence of trace dynamics. We generalised the theory to include gravitation. We start by assuming a Riemannian space-time  manifold with a spin structure,  inhabited by relativistic point particles. Thus, there exist on the manifold a metric, and the standard Dirac operator. No gravitational field equations are assumed. It is known from results in geometry that the Dirac operator and its square capture the information about the metric and curvature in a spectral manner \cite{Rovelli, Landi1999}. This property of the Dirac operator now plays a crucial role in arriving a matrix dynamical description of gravity.

In the spirit of trace dynamics, every space-time point [and its overlying metric]  is raised to the status of a (bosonic) matrix/operator. Each such matrix acts as a configuration variable, and comes with its own Dirac operator as a conjugate momentum variable. Moreover, each relativistic point particle is also raised to the status of a (fermionic) matrix. One does not treat the fermionic matter matrix and the bosonic space-time geometry it produces as segregated physical entities. Rather, they are respectively the fermionic [odd-grade Grassmann] and bosonic [even grade Grassmann] parts of  a Grassmann-valued matrix, dubbed an `atom' of space-time-matter [STM] or an `aikyon'. 

At the Planck scale, nature is assumed to be inhabited by enormously many such aikyons, which are operators in a Hilbert space, obeying a matrix dynamics. From here, the low energy world - quantum field theory as well as space-time and laws of general relativity - are emergent. It is significant that Einstein field equations with matter sources naturally emerge from the matrix dynamics - they are not put in by hand a priori - and are hence a prediction of the Planck scale matrix dynamics. The dynamics at the Planck scale is constructed ab initio.

Because space-time points have been raised to the status of matrices, we are in the realm of non-commutative geometry. Each aikyon obeys a non-commutative geometry, with the concept of distance and curvature being captured by its associated Dirac operator. Moreover, although classical space-time is lost, there emerges a new concept of time, intrinsic to a non-commutative geometry \cite{Connes2000},  and which we have named Connes time. We wrote an action principle principle for an aikyon evolving in Connes time; with the total action for all aikyons being the sum of their individual actions. From here, following the principles of trace dynamics, we derived the Lagrange equations of motion. Furthermore, at energies below Planck scale, there emerges after coarse-graining [so long as the imaginary part of the Hamiltonian is ignorable], the sought for formulation of quantum theory without classical space-time, the role of time now being played by Connes time, whereas there is no physical space yet. This is also a quantum theory of gravity, with the aikyon's configuration variables and momenta obeying quantum commutation relations and the Heisenberg equations of motion. There is also an equivalent Schrodinger picture. We note that this quantum gravity theory operates below the Planck scale, and is applicable whenever we want to find the quantum gravitational effects of a quantum system, without making reference to a background space-time. For instance, if we were to ask for the gravitational effect of the electron during a double slit experiment. This quantum gravity theory is an appropriate equivalent of quantum general relativity, that now also comes with a concept of time evolution, the one given by Connes time.

In this emergent quantum gravity, if a sufficiently many aikyons get entangled with each other, the imaginary part of the net Hamiltonian becomes significant. This results in a rapid breakdown of superposition, and spontaneous localisation results. This leads to the emergence of a space-time manifold, the one that was there before we raised space-time points to operators. Hence, spontaneous localisation is the reverse of the process of raising space-time geometry from Riemannian geometry to a non-commutative geometry. The emergent classical macroscopic bodies obey the laws of Einstein's general relativity. The overall scenario is described in the figure below, and in its accompanying caption [borrowed from \cite{maithresh2019b}].
\begin{figure}[!htb]
        \center{\includegraphics[width=\textwidth]
        {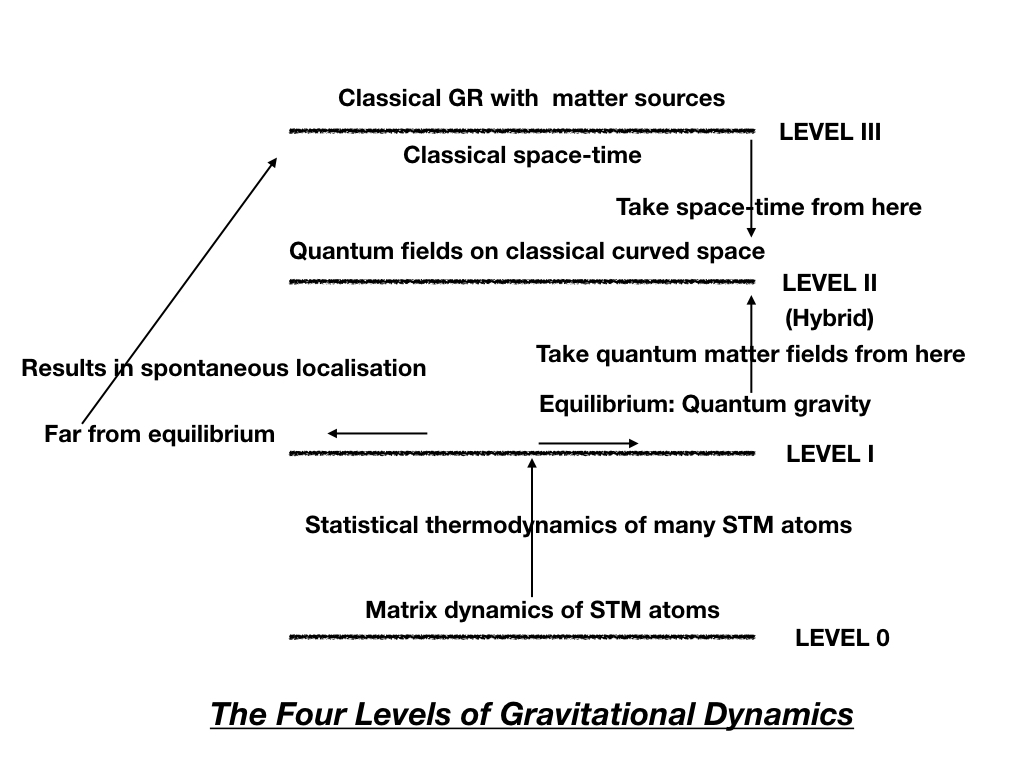}}
        \caption{\label{fig:my-label} The four levels of gravitational dynamics. In this bottom-up theory, the fundamental Level 0 describes the `classical' matrix dynamics of atoms of space-time-matter (STM). This level operates at the Planck scale. Statistical thermodynamics of these atoms brings us below Planck scale, to Level I: the emergent equilibrium theory is quantum gravity. Far from equilibrium, rapid spontaneous localisation results  in Level III: emergence of  classical space-time, obeying classical general relativity with matter sources. Level II is a hybrid level built by taking classical space-time from Level III and quantum matter fields from Level I, while neglecting the quantum gravitation of Level I. Strictly speaking, all quantum field dynamics takes place at Level I, but we approximate that to Level II. From \cite{maithresh2019b}.}
      \end{figure}

Given this backdrop of a classical universe, there is more than one way available now, for describing the dynamics of those degrees of freedom which have not undergone spontaneous localisation. At the Planck scale, their original aikyon matrix dynamics continues to hold. At energies below Planck scale, they are to be described as the emergent quantum gravity mentioned two paragraphs above. Or, as an approximation, and this is what quantum field theory is, their gravitational part can be neglected, and the [quantum] matter fields can be described as a quantum field theory on the background space-time. One way to arrive at this is to carry out trace dynamics for the un-localised degrees if freedom, on the space-time generated by macroscopic bodies. It is important to emphasise once again that the usage of space-time generated by external objects, to describe quantum dynamics, is a severe [though successful] approximation. An approximation which gives rise to the quantum non-locality puzzle: there is no such puzzle in the underlying matrix dynamics of the aikyons. 

This article gives an overview of spontaneous quantum gravity, following the path along which the theory was developed. Hence we start by reviewing spontaneous localisation - the phenomenological Ghirardi-Rimini-Weber theory which explains the absence of macroscopic quantum superpositions. Then follows Adler's theory of trace dynamics, which helps understand the origin of spontaneous localisation. In our search for a formulation of quantum theory without classical time, we realised this could be achieved by incorporating gravity into trace dynamics. This is described next.

The main sections are followed by a Track 2 section which give some mathematical details relevant to the preceding section. The reader who would like a quick overview of the theory can skip Track 2 sections, and read the rest of the paper without loss of continuity. An elementary overview is also available in \cite{QTST:2017, Singhsqg}.

\section{Spontaneous Localisation}

Text books on quantum mechanics often state that classical mechanics is obtained as the $\hbar\rightarrow 0$ limit of quantum mechanics. (In this limit, the Schr\"{o}dinger equation goes over to the classical Hamilton-Jacobi equation). However, such a statement hides an assumption: it is implicitly assumed, based on what we observe,  that position superpositions are absent in the classical world. In other words, even as the $\hbar\rightarrow 0$ limit is taken, a classical object could be in two or more locations at start of evolution (as allowed by quantum mechanics), and the Hamilton-Jacobi evolution would then imply that a classical particle would simultaneously evolve along a collection of trajectories; one trajectory per every initial location. The fact that such classical motion is never seen needs explaining, and is also the essence of the quantum measurement problem. That is, upon measurement, a pointer is never in more than one position at the  same time (unlike what the Schr\"{o}dinger equation predicts for the pointer) and the entangled state of the pointer and the measured quantum system collapse to one or the other classical outcomes. There is no universally accepted explanation as to why this should happen during the quantum-classical transition.

Here, it is important to emphasize that there is an intermediate regime between the microscopic and macroscopic, where quantum mechanics has not been experimentally tested. Simply because experiments in this intermediate regime are extremely challenging technologically, although important progress is taking place now [see e.g. the link tequantum.eu for the TEQ experiment]. This is brought out by the diagram below.
\begin{figure}[H]
	\centering
	\includegraphics[width=1.1\linewidth]{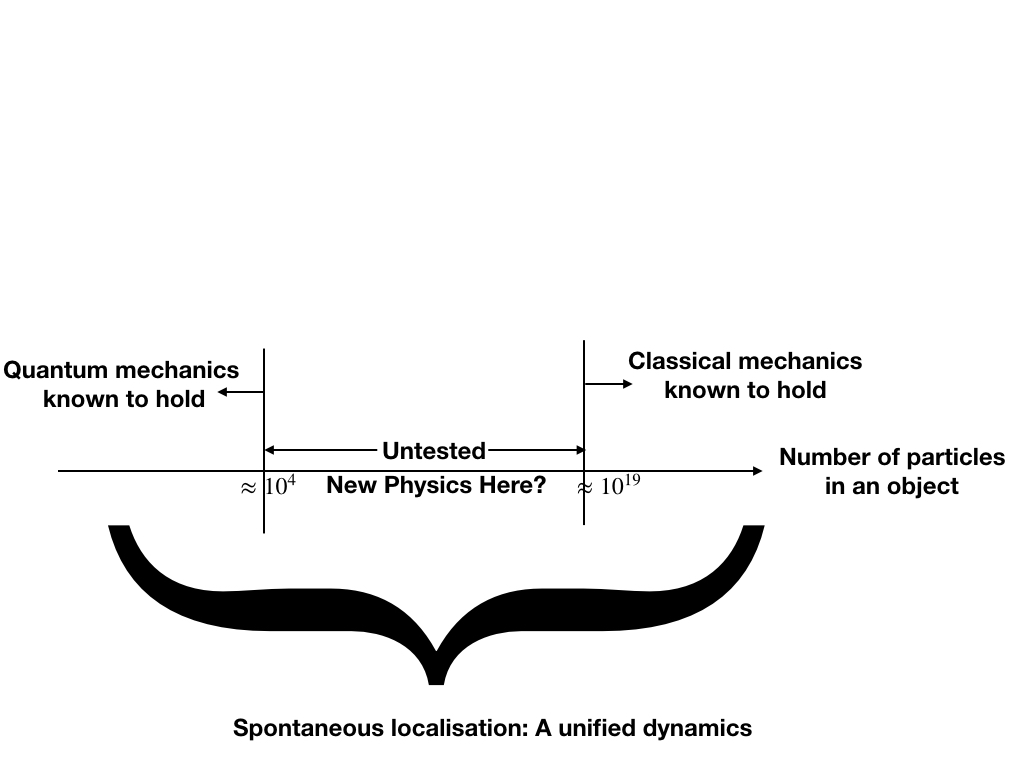}
	\caption{Tested and untested regimes of dynamics}
\end{figure}
The largest objects for which the principle of quantum linear superposition has been tested are made up of about twenty-five thousand nucleons (molecular interferometry). The smallest objects for which classical mechanics has been tested are made up of about $10^{19}$ nucleons (or an order or two less, in magnitude). There is thus an enormous desert of fifteen orders in magnitude, which is untested. New physics can arise here, in such a way that the new theory agrees with quantum mechanics for small objects, and with classical mechanics for large objects. However, the new theory ensures that during passage through this desert, the principle of quantum linear superposition breaks down dynamically. Such a breakdown is not ruled out by experiments, nor is it prohibited by the extraordinary success of quantum field theory.

The Ghirardi-Rimini-Weber-Pearle theory of spontaneous localisation, proposed first during 1970s and 1980s, achieves just that, providing a unified description of quantum and classical dynamics. The basic idea behind the theory is extremely simple, and beautiful. Recall that, according to the Schr\"{o}dinger equation, a quantum superposition lasts forever. Once a quantum system has been prepared in a superposition of, say two position eigenstates, it will evolve unitarily and stay in that superposed state for an infinite time. But clearly it is unphysical to talk of infinite time. Instead, GRW proposed, let us assume that the superposition of a nucleon in different position states lasts, on the average, as long as the age of the universe $\sim 10^{17}$ s. And then, it spontaneously and randomly collapses to one of those many eigenstates, with a probability given by the Born rule. Superposition is lost spontaneously \cite{Pearle:76,Pearle:89,Ghirardi2:90,Ghirardi:86,Bassi:03,RMP:2012,Mauro2019}.

This little change in the dynamics is enough to solve the quantum measurement problem, and to explain the absence of position superpositions in the macroscopic world. This is a consequence of quantum entanglement. Consider a bound macroscopic object [e.g. a chair]  whose atoms, all put together, have $N$ nucleons. Trying to create a superposition of the chair in two states, say chair here + chair there, amounts to creating an entangled state of the $N$ nucleons. It is easy to show that such an entangled state will spontaneously collapse in a time $T/N$ to one of the two position eigenstates, where $T$ is the spontaneous collapse mean life-time of one nucleon. The GRW theory assumes $T\sim 10^{17}$ s. If we take $N\sim 10^{23}$, the superposition will collapse in a millionth of a second. Thus superpositions are not absent in the macroscopic world; rather they are extremely short-lived. On the other hand, in the microscopic quantum world, superpositions last for a very long time (instead of lasting forever). In this way, spontaneous collapse theories provide a unified description of quantum and classical dynamics. Precisely in the untested region in the above diagram, the superposition life-time is neither too large nor too small, and differences from quantum and classical dynamics show up. Now when one takes the $\hbar\rightarrow 0$ limit, that procedure also destroys superposition, because an extra parameter is at play: the lifetime of the superposition.

The GRW theory can be cast into a precise mathematical formulation, by expressing it as a stochastic non-linear modification of the Schr\"{o}dinger equation. The non-unitary, non-linear part ensures breakdown of superposition, and its stochastic nature ensures randomness in collapse outcomes. Moreover the non-linearity is so constructed that evolution preserves norm, despite the non-unitarity. This ensures that the Born probability rule is reproduced. Also, a condition is imposed that the non-linearity should not lead to super-luminal signalling.

Whether or not dynamical collapse theories are correct can only be decided by experiment, and various ongoing experiments are pushing up the bound on the collapse time $T$. The current status is that we have two theories - quantum mechanics, and spontaneous localisation - which are both consistent with every experiment done to date. However only one of them can be correct, and that will be decided by further experiments.

Nonetheless, it is only fair to say that the theoretical structure of spontaneous collapse models has some shortcomings - overcoming these would only make these models more convincing. These models are ad hoc and phenomenological in nature, having been designed with the express purpose of solving the measurement problem, and explaining the quantum-classical transition. What is the fundamental origin of spontaneous collapse; what causes it? What is that stochastic noise which interacts with a quantum system, and introduces non-unitarity in its evolution? Why should norm be preserved despite the introduction of an external noise source? Its rather unusual in physics for stochastic effects to impact 
on a fundamental equation such as the Schr\"{o}dinger equation. Furthermore, collapse models are non-relativistic. Generalising them to a relativistic quantum-field theoretic version has remained  an unsolved problem, despite many serious efforts. All these shortcomings need to be overcome, to make spontaneous localisation into a robust physical theory that merges well with already known physical theories, such as relativistic quantum field theory. Recent developments address these issues, with rather dramatic implications, as we describe in the subsequent sections.

\section{Spontaneous Localisation: Track 2}

The Ghirardi-Rimini-Weber-Pearle  theory \cite{Ghirardi:86, Ghirardi2:90} proposes the following two postulates for dynamics in non-relativistic quantum mechanics:

1. Given the wave function $\psi ({\bf x_1}, {\bf x_2}, ..., {\bf x_N})$ of an $N$ particle quantum system in Hilbert space, the $n$-th particle undergoes  a `spontaneous localisatiion' to a random spatial position ${\bf x}$ as defined by the following so-called jump operator:
\begin{eqnarray}
{\psi_{t}({\bf x}_{1}, {\bf x}_{2}, \ldots {\bf x}_{N}) \quad
\longrightarrow \quad} 
 \frac{L_{n}({\bf x}) \psi_{t}({\bf x}_{1},
{\bf x}_{2}, \ldots {\bf x}_{N})}{\|L_{n}({\bf x}) \psi_{t}({\bf
x}_{1}, {\bf x}_{2}, \ldots {\bf x}_{N})\|}
\end{eqnarray}

This jump operator $L_{n}({\bf x})$ is a linear operator defined to be the normalised Gaussian:
\begin{equation}
L_{n}({\bf x}) =
\frac{1}{(\pi r_C^2)^{3/4}} e^{- ({\bf
\hat q}_{n} - {\bf x})^2/2r_C^2}
\end{equation}
${\bf \hat q}_{n}$ is the position operator for the $n$-th particle of the system and the random variable ${\bf x}$ is the spatial position to which the jump takes place. $r_C$, the width of the Gaussian, is a new constant of nature.

The probability density for the $n$-th particle to jump to the position
${\bf x}$ is assumed to be given by:
\begin{equation}
p_{n}({\bf x}) \quad \equiv \quad \|L_{n}({\bf x}) \psi_{t}({\bf
x}_{1}, {\bf x}_{2}, \ldots {\bf x}_{N})\|^2
\end{equation}
Also, it is assumed in the GRWP theory that the jumps are distributed in time as
a Poissonian process with frequency $\lambda_{\text{\tiny GRW}}$. This is the second
constant in this model.

2. Between
two consecutive jumps, the state vector evolves according to the
standard unitary Schr\"odinger equation.

These two postulates together provide a unified description of microscopic and macroscopic dynamics, and also an elegant solution to the quantum measurement problem (for reviews see e.g. \cite{Bassi:03, RMP:2012}).

The essential physics of spontaneous localisation can also  be described by a simple continuum model, known as QMUPL (Quantum Mechanics with Universal Position Localization), whose dynamics is given by the following stochastic nonlinear Schr\"{o}dinger equation
\begin{equation} \label{eq:qmupl1}
d \psi_t  =  \left[ -\frac{i}{\hbar} H dt + \sqrt{\lambda} (q - \langle q \rangle_t) dW_t  
 - \frac{\lambda}{2} (q - \langle q \rangle_t)^2 dt \right] \psi_t,
\end{equation}
where $q$ is the position operator of the particle, $\langle q \rangle_t \equiv \langle \psi_t | q | \psi_t \rangle$ is the quantum expectation, and $W_t$ is a standard Wiener process which encodes the stochastic effect. Evidently, the stochastic term is non-linear and also non-unitary. The collapse constant $\lambda$ sets the strength of the collapse mechanics, and it is chosen proportional to the mass $m$ of the particle according to the formula:
$
\lambda = \frac{m}{m_0}\; \lambda_0,
$
where $m_0$ is the nucleon's mass and $\lambda_0$ measures the strength of collapse. If we take $\lambda_0\simeq 10^{-2}$ m$^{-2}$ sec$^{-1}$ the strength of the collapse model corresponds to the GRWP model in the appropriate limit.

The above dynamical equation can be used to prove position localization. Let us onsider for simplicity a free particle  $(H=p^2/2m)$ in the gaussian state (analysis can be generalized to other cases):
\begin{equation} \label{gsol}
\psi_{t}(x) = \makebox{exp}\left[ - a_{t} (x -
\overline{x}_{t})^2 + i \overline{k}_{t}x + \gamma_{t}\right].
\end{equation}
By substituting this in the stochastic equation it can be proved that the spreads in position and momentum
\begin{equation}
\sigma_{q}(t)  \equiv  \frac{1}{2}\sqrt{\frac{1}{a_{t}^{\makebox{\tiny
R}}}};\qquad
\sigma_{p}(t)  \equiv  \hbar\,\sqrt{\frac{(a_{t}^{\makebox{\tiny R}})^2 +
(a_{t}^{\makebox{\tiny I}})^2}{a_{t}^{\makebox{\tiny R}}}},
\end{equation}
do not increase indefinitely but reach asymptotic values given by 
\begin{equation} \label{aval1}
\sigma_{q}(\infty) = \sqrt{\frac{\hbar}{m\omega}} \simeq
\left( 10^{-15} \sqrt{\frac{\makebox{kg}}{m}}\; \right)\,
\makebox{m}, \qquad
\sigma_{p}(\infty) = \sqrt{\frac{\hbar m\omega}{2}}
\simeq \left( 10^{-19} \sqrt{\frac{m}{\makebox{kg}}}\,
\right)\, \frac{\makebox{kg m}}{\makebox{sec}},
\end{equation}
such that:
$
\sigma_{q}(\infty)\, \sigma_{p}(\infty) \; = \;
{\hbar}/{\sqrt{2}}
$
which corresponds to almost the minimum limit permitted by Heisenberg's
uncertainty relation. Here, $\omega \; = \; 2\,\sqrt{{\hbar \lambda_{0}}/{m_{0}}} \; \simeq
\; 10^{-5} \; \makebox{s$^{-1}$}.$

Clearly, the spread in position does not increase indefinitely, rather it stabilizes to a finite value, which is a compromise between the Schr\"odinger dynamics, which spreads the wave function out in space, and the collapse dynamics, which shrinks it in space. For microscopic systems, this value is still relatively large ($\sigma_{q}(\infty) \sim 1$m for an electron, and $\sim 1$mm for a $C_{60}$ molecule containing some 1000 nucleons), such as to guarantee that in all standard experiments---in particular, diffraction experiments---one observes interference effects. For macroscopic objects however, the spread is indeed very small ($\sigma_{q}(\infty) \sim 3 \times 10^{-14}$m, for a 1g object). So small that for all practical purposes the wave function behaves like a point-like system. This is how spontaneous  localisation models are able to accommodate both the ``wavy'' nature of quantum systems and the ``particle'' nature of classical objects, within one single dynamical framework. 

The same stochastic differential equation solves the quantum measurement problem and also explains the Born probability rule without any additional assumptions. For illustration, consider a two state microscopic quantum system ${\cal S}$ described by the initial state
\begin{equation}
c_{+}|+\rangle + c_{-} |-\rangle
\end{equation}
interacting with a measuring apparatus ${\cal A}$ described by the position of a pointer which is initially in a `ready' state $\phi_{0}$ and which measures some observable $O$, say spin, associated with the initial quantum state of ${\cal S}$. As we have seen above, the pointer being macroscopic [for definiteness assume its mass to be 1 gram], is localized in a gaussian state $\phi^{G}$, so that the initial composite state of the system and apparatus is given by
\begin{equation} \label{eq:10}
\Psi_{0} = \left[ c_{+} |+\rangle + c_{-} |+\rangle \right]
\otimes \phi^{\mathrm{G}}.
\end{equation}
According to the standard quantum theory,  interaction leads to the following  evolution:
\begin{equation} \label{eq:6}
{\left[ c_{+} |+\rangle + c_{-} |-\rangle \right] \otimes \phi^{G}} \qquad
 \mapsto  \qquad  c_{+} |+\rangle \otimes \phi_{+}  + c_{-} |-\rangle
\otimes \phi_{-},
\end{equation}
where $\phi_{+}$ and $\phi_{-}$  are the final  pointer states corresponding to the system being in the collapsed state $|+\rangle$ or $|-\rangle$ respectively. While quantum theory explains the transition from the entangled state (\ref{eq:6}) to one of the collapsed alternatives by invoking a new interpretation or reformulation, the same is achieved dynamically by the stochastic nonlinear theory given by (\ref{eq:qmupl1}).

 It can be proved from (\ref{eq:qmupl1}) that the initial state (\ref{eq:10}) evolves, at late times, according to
\begin{equation}
  \label{eq:30}
  \psi_{t} = \frac{|+\rangle \otimes \phi_{+}
  + \epsilon_{t}|-\rangle \otimes \phi_{-}}{\sqrt{1+ \epsilon_t^2}}.
\end{equation}
The evolution of the stochastic quantity $\epsilon_t$ is determined dynamically by the stochastic equation: it either goes to $\epsilon_t \ll 1$, with a probability  $|c_{+}|^2$, or to $\epsilon_t \gg 1$, with a probability 
$|c_{-}|^2$. In the former case, one can say with great accuracy that the state vector has `collapsed' to the definite outcome $|+\rangle \otimes \phi_{+}$ with a probability $|c_{+}|^2$. Similarly, in the latter case one concludes that the state vector has collapsed to $|-\rangle \otimes \phi_{-}$ with a probability $|c_{-}|^2$.
This is how collapse during a quantum measurement is explained dynamically, and random outcomes over repeated measurements are shown to occur in accordance with the Born probability rule. The time-scale over which $\epsilon_t$ reaches its asymptotic value and the collapse occurs can also be computed dynamically. In the present example, for a pointer mass of one gram, the collapse time turns out to be about $10^{-4}$ s.

Furthermore, we can also understand how the modified stochastic dynamics causes the outcome of a diffraction experiment in matter wave-interferometry to be different from that in quantum theory. Starting from the fundamental equation (\ref{eq:qmupl1}) it can be shown that the statistical operator $\rho_t = \mathbb{E}[|\psi_t\rangle\langle\psi_t|]$ for a system of $N$ identical particles evolves as
\begin{equation}
\rho_t(x,y) = \rho_0(x,y) e^{- \lambda N (x-y)^2 t/2}.
\end{equation}
Experiments look for a decay in the density matrix by increasing the number of  particles $N$ in an object, by increasing the slit separation $|x-y|$, and by increasing the time of travel $t$ from the grating to the collecting surface.  The confirmed detection of an interference pattern sets an upper bound on $\lambda$. The absence of an interference pattern would confirm the GRWP theory and determine a specific value for $\lambda$ (provided all sources of noise such as decoherence can be  ruled out.)

A detailed review of the spontaneous localisation model and its experimental tests and possible underlying theories can be found in ~\cite{RMP:2012}. The GRW theory motivated us to propose that space-time itself arises from the collapse of the wave-function \cite{Singh:2019}. If no GRW collapses were to take place, everything in the universe would be quantum, and there would then be no classical space-time either. Hence there ought to be a way of formulating the GRW theory without reference to space-time. This means that spontaneous localisation must give rise to space-time along with giving rise to classical behaviour of macroscopic objects. Since space-time is expected to emerge from an underlying quantum theory of gravity, spontaneous localisation must also be emergent from quantum gravity. This happens naturally in our matrix dynamics based quantum gravity: hence the name Spontaneous Quantum Gravity.

\section{The theory of trace dynamics}

Adler's theory of trace dynamics \cite{Adler:94, Adler-Millard:1996, Adler:04} is built on the guiding principle that quantum theory, being more fundamental than classical mechanics, should be constructed ab initio from first principles in a bottom-up fashion. Rather than our having to arrive at quantum (field) theory by quantizing the theory's own limit, viz. classical dynamics. Thus, we do not arrive at special relativity by relativizing Newtonian mechanics. Nor do we arrive at general relativity by relativizing Newton's law of gravitation. The two relativity theories are built from their own new concepts and symmetry principles (universal constancy of speed of light, and interpretation of gravitation as space-time curvature). Newtonian dynamics then naturally follows as the non-relativistic limit of the relativity theory. 

Trace dynamics is the classical matrix dynamics of matrices (equivalently operators) on a Minkowski space-time. A matrix describes an elementary particle or a field; the idea being that instead of using $c$-numbers or real numbers to describe these entities, one uses matrices. The consequences are far-reaching. Each matter or field degree of freedom is described by an operator degree of freedom, labelled say ${\bf q}$ (configuration variable). ${\bf q}$ is a function of time if it describes a particle, and of space-time if it describes a field. Thought of as a matrix, ${\bf q}$ is made of Grassmann numbers as its elements. Grassmann numbers anti-commute with each other. Such a matrix can always be written as a sum of a `bosonic' matrix and a `fermionic' matrix. A  bosonic matrix is even grade Grassmann (matrix elements made of product of even number of Grassmann elements, so that they commute with each other), and a fermionic matrix is odd grade Grassmann (matrix elements made of product of odd number of Grassmann elements, so that these anti-commute amongst themselves). The nomenclature is natural, as bosonic / fermionic matrices are indeed used to describe bosonic / fermionic fields, in particle physics. An operator polynomial made from ${\bf q}$s and its time derivatives is used to construct a Lagrangian, by taking the matrix trace of this polynomial - the trace Lagrangian, as it is referred to. As in classical dynamics, time integral of the trace Lagrangian defines the action. Equations of motion are derived  by extremizing the action, while varying with respect to the ${\bf q}$s (using the trace derivative). One arrives at Lagrange's equations of motion, from which a Hamiltonian dynamics can also be constructed. All configuration variables as well as their canonically conjugate momenta obey arbitrary commutation relations with each other, which inevitably evolve with time, consistently with the equations of motion. Thus this is a classical dynamics because it follows from variation of the action / Lagrangian, but in a sense it is even  `more quantum' than quantum mechanics because the commutation relations are arbitrary (not fixed like the quantum commutation relations). There is no Planck's constant yet - $\hbar$ is emergent in this theory.

In spirit, trace dynamics (a classical matrix dynamics) resembles matrix models which have been studied in the past, including in the context of string theory. The central difference between matrix models and trace dynamics is that one does not quantise trace dynamics. On the contrary, quantum (field) theory is derived from trace dynamics as the statistical thermodynamics of a large number of $\bf q$ matrices, by coarse-graining their evolution in operator phase space. Thus, trace dynamics is assumed to hold at the Planck scale, and one would like to examine what is the dynamics much below the Planck scale. This is where statistical mechanics comes in. The system point is assumed to visit all allowed states in the phase space, so that long time averages may equal ensemble averages. A probability distribution is defined in phase space, using a suitable measure, and the equilibrium distribution is determined by maximising the von Neumann entropy.

On the physical front, what distinguishes trace dynamics from Newtonian mechanics is the existence of a remarkable conserved charge, which results from a global unitary invariance of the trace Hamiltonian. This charge, known as the Adler-Millard charge, is given by the sum over all bosonic degrees of freedom of their respective commutators $[q,p]$, minus the sum over all the fermionic degrees of freedom, of their respective anti-commutators $\{q,p\}$. Each of these commutators has dimensions of action, and is by itself time-dependent. Yet the Adler-Millard charge defined from them is conserved. It turns out that at equilibrium, this charge is equipartitioned over all the degrees of freedom - the equipartitioned value is identified with Planck's constant $\hbar$. It is shown that at equilibrium, the ensemble averages of the canonical degrees of freedom obey the Heisenberg equations of motion. This is how quantum (field) theory is derived from first principles, by starting from a well-defined matrix dynamics. An equivalent Schr\"{o}dinger functional picture can also be constructed, as in quantum field theory.

The next significant move is to recognise that there always are statistical fluctuations around equilibrium, such as those which are responsible for Brownian motion. Such fluctuations modify the evolution equations of quantum (field) theory. In principle, the corrections can include a non-self adjoint component as well, which causes the appearance of anti-self adjoint corrections to the Hamiltonian.

Adler considered the role of these corrections in the context of the non-relativistic Schr\"{o}dinger equation, for matter (fermionic) degrees of freedom. This amounts to adding a stochastic correction (including an anti-self-adjoint part) to  the matter Hamiltonian. The structure now is pretty much as in collapse models. Assuming, as in collapse models, that norm is preserved (despite non-unitary evolution), and that superluminal signalling 
is not allowed, one arrives at a stochastic non-linear Schr\"{o}dinger equation with the same structure as a collapse model. The theory of trace dynamics can hence explain the origin of spontaneous localisation - the latter is no longer an ad hoc proposal.

In our recent work we have addressed the unresolved issues in trace dynamics. Amongst these are the following. Trace dynamics is formulated at the Planck scale, but it assumes the space-time background to be Minkowski. It would be more natural to allow for a quantum behaviour of space-time, and to incorporate gravity, albeit not as classical gravity, but as operator gravity. We solve this problem, by bringing in the description of space-time structure from Connes' non-commutative geometry programme. Secondly, in trace dynamics, only a non-relativistic theory of spontaneous collapse is arrived at. By bringing in gravity, we construct a relativistic theory of spontaneous localisation. Thirdly, we explain why only the fermionic (matter) degrees of freedom undergo spontaneous collapse, whereas bosonic degrees (the gravitational field say) do not. And we also explain why the norm of the evolving state vector must be preserved, despite the presence of anti-self-adjoint corrections to the Hamiltonian.

\section{The theory of trace dynamics: Track 2}
As noted above, race dynamics [TD] derives quantum (field) theory and spontaneous localisation from an underlying (pre-quantum) matrix dynamics. It is the dynamics of matrix models which obey a global unitary invariance, operating at the Planck scale; However, as an approximation, space-time is assumed to be Minkowski space-time, and gravity is not included in the theory. Suppose we take classical dynamics [either Newtonian mechanics or special relativity] as the starting point, and instead of describing a material point particle by a real number, we describe it by a matrix (equivalently, operator). This is the essence of trace dynamics. For a particle $q$, now described by a matrix ${\bf q}$, the action is changed as in this example:
\begin{equation}
S = \int dt \ [\dot{q}^2 - q^2] \quad  \longrightarrow \int dt \ Tr\  [\dot{\bf q}^2 - {\bf q}^2]
\end{equation}
After replacing the configuration variable $q$ by a matrix, the scalar Lagrangian is constructed by taking a matrix trace of the operator polynomial, and then a scalar action is constructed as usual,  by integrating the trace Lagrangian over time. A general trace Lagrangian ${\bf L}$ is a function of the various configuration variables ${\bf q}_i$ and their time derivatives $\dot{\bf q}_i$, and is made from the trace of an operator polynomial ${\cal L}$. This construction can be extended to field theory by raising the field value at each space-time point to a matrix, then constructing an operator polynomial, taking its trace to form a Lagrangian density, and integrating over four-volume to get the action [continuum limit].

Lagrange equations of motion are obtained by varying the action with respect to the operator ${\bf q}_i$. In order to vary the trace Lagrangian with respect to an operator, the notion of a trace derivative is introduced. The derivative of the trace Lagrangian ${\bf L}$  with respect to an operator $\cal O$ in the polynomial ${\cal L}$ is defined as follows:
\begin{equation}
\delta {\bf L }= Tr \frac{\delta{\bf L}}{\delta\cal{O}}\delta\cal{O}
\end{equation}
This so-called trace derivative is obtained by varying ${\bf L}$ with respect to ${\cal O}$ and then cyclically permuting ${\cal O}$ inside the trace, so that $\delta\cal O$ sits to the extreme right of the polynomial $\cal{L}$. 

It is assumed that the matrix elements are complex valued Grassmann numbers, which can be further sub-divided into even grade and odd grade Grassmann numbers. Any Grassmann matrix can be written as a sum of two matrices: the bosonic part (made of even grade elements) and the fermionic part (made of odd grade elements). Bosonic (fermionic) matrices describe bosonic (fermionic) fields, as in conventional quantum field theory. Thus, in trace dynamics there are both bosonic degrees of freedom ${\bf q}_B$ and fermionic degrees of freedom ${\bf q}_F$. Recently, we have proved that in our matrix dynamics there is a natural definition of spin, ad in the emergent quantum theory fermionic matrices have half-integer spin, and bosonic matrices have integral spin.

The Euler-Lagrange equations
\begin{equation}
\frac {d\;}{dt}\Big ( \frac{\delta {\bf L}}{\delta \dot {\bf q}_i}\Big ) - \Big ( \frac {\delta {\bf L}}{\delta {\bf q}_i}\Big ) = 0
\end{equation} 
are used to obtain the operator equations of motion,  and they also define the canonical momenta. The configuration variables and the momenta do not commute amongst each other, and the commutation relations are determined by the dynamics. This is what makes trace dynamics different from both classical dynamics as well as from quantum theory. Apart from the trace Hamiltonian, 
\begin{equation}
{\bf H}=\sum_i \mathrm{Tr}[p_{Fi}\dot{q}_{Fi}]+\sum_i \mathrm{Tr}[p_{Bi}\dot{q}_{Bi}]-\mathrm{Tr}\; \mathcal{L}
\end{equation}
there is another conserved charge of great importance; the Adler-Millard charge, denoted as $\tilde{C}$. This charge is a consequence of a global unitary invariance of the trace Lagrangian and the trace Hamiltonian. It is given by the expression
\begin{equation}
    \tilde{C} = \sum_{r\in B}[q_r,p_r] -\sum_{r\in F} \{q_r,p_r \} 
    \label{amc}
\end{equation}
We shall henceforth drop the bold notation from the canonical variables, it being understood that we deal with matrix/operator valued canonical variables. The Adler-Millard charge is unique to matrix dynamics, and plays a central role in emergence of quantum theory from trace dynamics. If the trace Hamiltonian is self-adjoint, then the Adler-Millard charge can be shown to be anti-self-adjoint. Were the trace Hamiltonian to have an anti-self-adjoint component, this conserved charge picks up a 
self-adjoint component - this will be important for us when we incorporate gravity in trace dynamics.

Hamilton's equations of motion are given by
\begin{align}
    \frac{\delta \textbf{H}}{\delta q_r} =-\dot{p}_r, \quad \quad \frac{\delta \textbf{H}}{\delta p_r} =\epsilon_r \dot{q}_r
\end{align}
where $\epsilon_r = 1(-1)$ when $q_r$ is bosonic(fermionic). 

The above dynamics is Lorentz invariant, and is  assumed to take place at the Planck energy scale. TD does not specify the form of the fundamental Lagrangian, though we will choose a particular form below when we incorporate gravity into TD. Since the physical systems that we observe and experiment with, operate at energy scales much lower than Planck scale and are not probed over Planck times, we ask the following question: What is the averaged description of trace dynamics, if we coarse grain (smear)  the trace dynamics over time intervals much larger than Planck times? We might imagine that there are extremely rapid variations in the canonical variables over Planck time scales, but there is a smoothed out dynamics at lower energies, where these rapid variations have been coarse-grained over. The methods of statistical thermodynamics are employed, treating the underlying dynamics as `microscopic' degrees of freedom, to show that the emergent 
coarse-grained dynamics  is relativistic quantum (field) theory.

One begins by constructing the  phase space of matrix dynamics, with (the real and imaginary parts of) each element $(q_{r})_{lm}$ of $q_r$ being a (pair of) independent degrees of freedom in the phase space, along with the matrix component (again real and imaginary part) $(p_i)_{im}$ of the corresponding momentum. We use the symbol $x$ to denote $q$ or $p$.  A measure $d\mu$ is defined in the phase space, as
\begin{equation}
(x_r)_{mn} = (x_r)^{0}_{mn} + i (x_r)^{1}_{mn}; \quad d\mu = \prod_A d\mu^{A}; \quad d\mu^{A}=\prod_{r,m,n} d(x_r)^{A}_{mn}
\end{equation}
where $A=0,1$ and the components $(x_r)^{A}_{mn}$ are real numbers. This measure is conserved during evolution, and obeys Liouville's theorem. Moreover, the measure is invariant under infinitesimal operator shifts $x_r\rightarrow x_r + \delta x_r$. 

A phase space probability density distribution  $\rho[\{x_r\}]$ is defined in the matrix element phase space. This determines the probability of finding the system point in some particular infinitesimal volume in phase space. A canonical ensemble is constructed for a sufficiently large number of identical systems, each of which start evolving from arbitrary initial conditions in the phase space. It is assumed that over time intervals much larger than Planck time, the accessible region of the phase space [i.e. the region allowed by  a conserved trace Hamiltonian and a conserved Adler-Millard charge] is uniformly populated, and hence that the long time average [the coarse-grained dynamics] can be determined from the ensemble average at any one given time.  This equilibrium dynamics is determined as usual,  by maximising the Boltzmann entropy
\begin{equation}
\frac{S_E}{k_B}=-\int d\mu \; \rho\; \ln\rho
\label{entro}
\end{equation}
This is done subject to the constraints that the ensemble-averaged trace Hamiltonian $\langle {\bf H} \rangle_{AV}$ and the ensemble averaged Adler-Millard charge $\langle {\bf \tilde{C}}\rangle_{AV}$ are conserved. These two constraints are imposed by introducing the Lagrange multipliers $\overline\tau$ and $\tilde{\lambda}$ respectively, where $\overline\tau$ is a real constant with dimensions of inverse mass, and $\overline\lambda$ an anti-self-adjoint  matrix with dimensions of inverse action.  

Hence the phase space density distribution $\rho$ depends, apart from the dynamical variables, on $\tilde{C}, \tilde{\lambda}, {\bf H},\overline \tau$ and can be written as $\rho(\tilde{C}, \tilde{\lambda}, {\bf H},\overline \tau)$. It can be further shown that the dependence on $\tilde{C}$ and $\tilde{\lambda}$ is of the form $Tr(\tilde{\lambda}\tilde{C})$, so we write $\rho = \rho(Tr[\tilde{\lambda}\tilde{C}],\overline\tau,{\bf H})$. It can be shown, subject to the plausible assumption that the ensemble does not favour any one state in the ensemble over the other, that the canonical ensemble average of the Adler-Millard charge takes the form
\begin{equation}
\langle \tilde{C} \rangle_{AV}  = i_{eff} \hbar; \qquad i_{eff} = i\ diag(1,-1,1,-1...,1,-1)
\end{equation}
where the real constant $\hbar$ is eventually identified with Planck's constant, subsequent to the emergence of quantum dynamics.

The equilibrium distribution is arrived at by maximising the function $-\cal{F}$ where
\begin{equation}
{\cal F} = \int d\mu\ \rho \log \rho + \theta \int d\mu\ \rho + \int d\mu\ \rho Tr \tilde{\lambda}\tilde{C} +\overline \tau \int d\mu\ \rho {\bf H}
\end{equation}
and gives the result
\begin{align}
    \rho = Z^{-1} \exp{(-\text{Tr} \Tilde{\lambda} \Tilde{C} -\tilde\tau \textbf{H})} \label{rho} \\
    Z = \int d \mu \; \exp{(-\text{Tr} \Tilde{\lambda}\Tilde{C} -\tilde\tau \textbf{H})} \label{z}
\end{align}
The entropy at equilibrium is given by the expression
\begin{equation}
   \frac{S_E}{k_B} =  \log{Z} - \text{Tr}\tilde{\lambda} \frac{\partial \log{Z}}{\partial \tilde{\lambda}} -\tilde \tau \frac{\partial \log{Z}}{\partial \tilde\tau}  \label{entropy}
\end{equation}

We ask: what is the mean dynamics obeyed by the variables $\langle x\rangle_{AV}$, averaged over the canonical ensemble, at energy scales below Planck scale? To answer this, one derives certain Ward identities, as is done for  functional integrals in quantum field theory, in analogy with the proof for the equipartition theorem in statistical mechanics. These identities are a consequence of the invariance of the phase space measure under constant shifts of the dynamical variables. Thus, in conventional statistical mechanics,  the equipartition theorem is a consequence of the vanishing of the integral of a total divergence:
\begin{equation}
0 = \int d\mu \ \frac{\partial [x_r \exp(-\ \beta H)]}{\partial x_s}
\end{equation}
In the statistical mechanics of trace dynamics, we have for a general operator $\cal{O}$, that its average over the canonical ensemble is unchanged when a dynamical variable is varied:
\begin{equation}
0 =  \int d\mu\ \delta_{x_r} (\rho{\cal O})
\end{equation}
One chooses ${\cal O}$ to be the operator $Tr\{\tilde{C}, i_{eff}\}W$ where $W$ is any bosonic polynomial function of the dynamical variables, and carries out the above variation, taking $\rho$ to be the equilibrium phase space density distribution function. Thus we have 
\begin{equation}
0 = \int d\mu \ \delta_{x_r}\ \left[\exp \left(-Tr \tilde{\lambda}\tilde{C} - \tilde\tau {\bf H} \right)\  Tr \{\tilde{C},i_{eff}\}W\right]
\end{equation}
A very important assumption is made, namely that $\tilde{\tau}$ is the Planck time scale, and that we are interested in the averaged dynamics over much larger time scales (equivalently much lower energies). Each dynamical variable $x_r$ is split into a `fast' varying part [which varies over Planck times] and a `slow' part which is constant over Planck times. Important conclusions then follow from the above Ward identity, by making different choices for $W$. When $W$ is chosen to be a dynamical variable $x_r$, standard quantum commutation relations for bosonic and fermionic degrees of freedom are shown to be obeyed by the averaged variables $\langle x_r\rangle_{AV}$. The constant $\hbar$ introduced above is identified with Planck's constant. If $W$ is identified with the operator polynomial $H$ whose trace is the trace Hamiltonian ${\bf H}$, the quantum Heisenberg equations of motion for the averaged dynamical variables are obtained. The underlying matrices of TD, within ensemble averages, obey properties analogous to quantum fields. The contact with quantum field theory is made as follows. There is a unique eigenvector $\psi_0$ whose corresponding eigenvalue is the lowest eigenvalue of $H$. This acts as the conventional vacuum state, and canonical ensemble averages are identified with Wightman functions in the emergent quantum field theory, for a given function S,
\begin{equation}
\psi^{\dagger}_0 \  \langle S \{x_r\}\rangle_{AV} \ \psi_0 = \langle vac | S\{X\}|vac\rangle
\end{equation}
where $X$ is a quantum field operator.
In this way, relativistic quantum
(field) theory is shown to arise as an emergent phenomenon, it being the low energy equilibrium approximation in the statistical thermodynamics of an underlying matrix dynamics. Once the Heisenberg equations of motion are known, one can also transform to the functional Schrodinger picture in the standard manner.

The theory of trace dynamics also provides a theoretical basis for the origin of the phenomenological theory of spontaneous localisation. As we have seen above, quantum dynamics is a mean dynamics arising from averaging over Planck time scales, and neglecting the fast component in the variation of the dynamical variables. Under certain circumstances, the fast component can become significant, in which case its impact on the coarse-grained dynamics can be modelled as stochastic fluctuations around equilibrium. Particularly crucial is that these fluctuations can make an ant-self-adjoint stochastic contribution to the quantum theory Hamiltonian. This is possible because the underlying trace Hamiltonian can have a small anti-self-adjoint part at the Planck scale, which could get amplified by entanglement between a very large number of particles. Precisely such a situation arises when gravity is included in trace dynamics, as we will see below.

Adler considers such a possibility for fermions in the theory, in the non-relativistic approximation to quantum field theory, where the anti-Hermitean fluctuating correction to the Hamiltonian is modelled by adding a stochastic function ${\cal K}(t)$ 
\begin{equation}
i\hbar \frac{\partial \Psi}{\partial t} = H\Psi + i{\cal K}(t)\Psi
\end{equation}
This modified equation however does not preserve norm of the state vector during evolution. If we do insist on norm-preservation, and transform to a new state vector whose norm is preserved, the resulting evolution equation is non-linear. It also makes the evolution non-unitary. If we also demand that the non-linear evolution should not lead to superluminal signalling, the form of the evolution becomes just the same as in spontaneous localisation models. Thus trace dynamics can in principle explain the quantum-to-classical transition, by taking into consideration the potential role of statistical fluctuations around equilibrium. The theory provides a common origin for quantum theory, as well as for spontaneous localisation, starting from an underlying matrix dynamics possessing a global unitary invariance.

Trace dynamics does not specify the fundamental Lagrangian for physical interactions. Also, it does not include gravity, although it operates at the Planck scale. The theory also leaves some important questions unanswered. For instance, what is the origin of the small anti-self-adjoint  component of the Hamiltonian at the Planck scale? Why does spontaneous localisation take place only for fermions, but not for bosons? Why should the norm of the state vector be preserved despite the presence of the 
anti-Hermitean fluctuations? In the next section, we demonstrate how to include gravity in trace dynamics, using the principles of Connes' non-commutative geometry programme - this leads us to a candidate quantum theory of gravity, for which we specify a Lagrangian. We also answer the open questions left unanswered by trace dynamics, as mentioned in the preceding lines.

\section{Incorporating gravity in trace dynamics}

We emphasize that our primary motive behind this approach to quantum gravity was not that of incorporating gravity in trace dynamics. Rather, our goal was to arrive at a formulation of quantum (field)  theory which does not refer to classical space-time. The realisation that such a formulation must exist is the single most important clue towards a quantum theory of gravity. Classical space-time is a consequence of the universe being dominated by classical macroscopic bodies. In the absence of such bodies (which in fact are a limiting case of quantum systems, thus forcing quantum theory to depend on its own limit) there will be no space-time, yet we should we able to describe quantum systems (without appealing to classical time). Thus, we do  not quantize space-time; rather we get rid of space-time from quantum theory - this leads to a falsifiable candidate quantum theory of gravity, which predicts spontaneous localisation.

Our underlying physical principle/symmetry is to demand that the laws of gravitation, and of the matter sources that describe them, are invariant under general coordinate transformations of non-commuting coordinates. This takes us to the domain of Connes' non-commutative geometry. This symmetry principle also has the flavour of trace dynamics, because the non-commuting coordinates are operators (equivalently matrices) which obey arbitrary commutation relations amongst them. 

Non-commutative geometry (NCG) provides a spectral view of gravitation and curvature, which again ties in well with trace dynamics \cite{Landi1999}. The relevant result for us, which we present here in a simplified manner, is the following. Given a Riemannian manifold describing (Euclidean) curved space-time, construct the standard Dirac operator on this space-time, and find its eigenvalues. The sum of the squares of these eigenvalues is equal, up to constants,  to the Einstein-Hilbert action on that space-time! That is, denoting the Dirac operator by $D_B$, and trace of its square  by $Tr[L_P^2 D_B^2]$, we have
\begin{equation}
Tr [L_P^2 D_B^2] \sim \frac{1}{L_P^2}\int d^4x \sqrt{g} R
\end{equation}
Next, if we make the algebra of coordinates non-commutative, we no longer have the original space-time manifold, but we still have the spectral description of its curvature, as on the left hand side of the above equation. It is hence assumed that $Tr [L_P^2 D_B^2]$ 
describes curvature of the non-commutative geometry. This celebrated spectral action, as it is called in non-commutative geometry, points to a deep connection between the Dirac operator and gravitation, and plays a crucial role in our quantum theory of gravity.

The second relevant and extremely significant result from NCG is the existence of a fundamental time parameter, which is there only in the non-commutative case, and absent in ordinary commutative geometry. This is a consequence of the so-called Tomita-Takesaki theory, and the `co-cycle Radon-Nikodym' theorem. For us it suffices to note that there are a one-parameter family of  inner automorphisms of the non-commutative algebra,  which map elements of the algebra to other elements of the algebra; this being equivalent to a time translation. As Connes puts it, `non-commutative measure spaces evolve in time'. We call this Connes time, and denote it by $\tau$. When ordinary space-time is lost because of non-commutativity, Connes time emerges, and helps us to  
formulate quantum theory without classical time.

Because the spectral action does not depend on the existence of a space-time manifold,
(and yet links to classical gravitation), it has the right properties for inclusion in trace dynamics. But with a twist. For a physicist, for something to be an action, it should be the time integral of a Lagrangian. Here, Connes time comes to our rescue, noting also that the `spectral action' $Tr[L_P^2 D_B^2]$ is more in tune with what we would call a trace Lagrangian in trace dynamics. Furthermore, a trace Lagrangian should be an operator polynomial made from a configuration variable and its time derivatives. This motivated us to define a bosonic configuration variable $q_B$ as follows: $D_B \equiv (1/Lc) dq_B/d\tau$, and hence a trace Lagrangian and a trace action:
\begin{equation}
S_B = \frac{1}{\tau_{Pl}}\int d\tau\; Tr \bigg[\frac{L_P^2}{L^2 c^2} \left(\frac{dq_B}{d\tau}\right)^2\bigg]
\end{equation}
Here, $L$ is a length scale associated with $q_B$, and $q_B$ is related to gravitation through the eigenvalues of $D_B$. This is how we have used NCG to incorporate gravity into trace dynamics \cite{maithresh2019}.

This Lagrangian also helps us arrive at a formulation of quantum theory without classical time. To progress in that direction we must now introduce matter (fermions) and relate matter to gravity in trace dynamics, analogous to the spirit of classical general relativity. Keeping in mind that this matter ought to be quantum in nature, it is perfectly reasonable to assume (since quantum systems are not localised in space) that we should no longer make a distinction between fermionic matter and the gravitation it produces. To this end we introduced the concept of an atom of space-time-matter (STM), denoted by operator $q$, which is split into its bosonic and fermionic parts as $q=q_B + q_F$, with $q_B$ defined as above, and $q_F$ the matter (fermionic) part. The constraint on $q_F$ is that it should be possible to identify it, upon the emergence of classical space-time, as the matter degree of freedom in quantum theory, and in general relativity. Thus the STM atom carries around its own (non-commutative) geometry. An STM atom is an elementary particle plus its own space-time geometry. If we ask what is the gravitational field of an electron, we would describe the electron and its gravity together as an STM atom. At the Planck scale, the universe is populated by enormously many STM atoms, each described by its own $q$-operator, whose dynamics is described in the Hilbert space via evolution in Connes time. The fundamental action principle for an STM atom is
\begin{equation}
\frac{L_P}{c} \frac{S}{C_0}  =  \frac{1}{2} \int d\tau \; Tr \bigg[\frac{L_P^2}{L^2c^2}\; (\dot{q}_B +\beta_1 \dot{q}_F)\;(\dot{q}_B +\beta_2 \dot{q}_F) \bigg]
\end{equation}
Here $\beta_1$ and $\beta_2$ are two constant fermionic matrices whose properties remain to be determined. This action looks similar to the action for a free point particle in classical mechanics, except that now the configuration variable does not describe just matter, but also its gravity. It is interesting that the particle description (as opposed to the description via fields) comes back in full force in our matrix dynamics. This is understandable, because classical space-time is lost, and it would not be meaningful to talk of fields when physical three-space is not there. 

The equations of motion and their solutions  obtained from this action are highly instructive. These are:
\begin{align}
    2\dot{q}_B +(\beta_1+\beta_2)\dot{q}_F = c_1 \\
    \dot{q}_B(\beta_1+\beta_2)+ \beta_1 \dot{q}_F \beta_2 +\beta_2 \dot{q}_F \beta_1 = c_2
\end{align}
where $c_1$ and $c_2$ are constant bosonic and fermionic matrices, respectively. These two equations are the (matrix dynamics) precursors of the Einstein-Dirac equation and the Schr\"{o}dinger-Newton equation [matter tells space-time how to curve; space-time 
tells matter how to move]. 

There is one such action term for every STM atom. It is not as if all the STM atoms together produce gravitation of the universe; rather classical space-time emerges after the fermionic parts of many entangled atoms undergo spontaneous localisation. This way, the material bodies of the universe are formed, and formed concurrently with the emergence of space-time.

\section{Incorporating gravity in trace dynamics: Track 2}
In the Introduction section, we have argued that there ought to exist an equivalent reformulation of quantum (field) theory which does not refer to classical space-time. One possible way to arrive at such a reformulation is to raise space-time points to the status of non-commuting matrices/operators, in the spirit of what was done in trace dynamics above for material particles / matter fields. 
Non-commutative geometry [NCG] allows for such a possibility for space-time and its geometry. In other words, Connes' NCG
program does for space-time what trace dynamics does for matter fields. We propose to put non-commutative geometry together with trace dynamics, and propose a matrix dynamics for matter Dirac fermions and the (non-commutative) space-time geometry  produced by them. This new theory operates at Planck time/energy scales, just as TD does. The statistical thermodynamics of this new theory - i.e. coarse-graining over times larger than Planck time, provides us with a candidate quantum theory of gravity, which is also the sought for quantum (field) theory without classical time \cite{maithresh2019}.

Hereon we will assume a Euclidean space-time, and Euclidean general relativity. The case of Lorentzian space-times still remains to be dealt with. In NCG \cite{Connes2000}, the definition of a spectral action derives from the spectral definition of infinitesimal distance using the distance operator $d\hat{s}$. This operator is  related to the Dirac operator $D$ as $d\hat{s}=D^{-1}$, thus providing a definition of distance - equivalent to the standard definition of distance [in terms  of the metric] - as and when a Riemannian geometry and a manifold exists. This spectral definition of distance continues to hold also when an underlying space-time manifold is absent, as for instance when the algebra of coordinates does not commute [transition to non-commutative geometry]. 

Next, the integral $\fint T$ of a first order infinitesimal in operator $T$ is defined to be the coefficient of the logarithmic divergence in the Trace of T \cite{Connes2000}. We may visualise the integral of an operator as if it were  the sum of its eigenvalues. The spectral action relating to gravity $S$ is defined as the slash integral
\begin{equation} 
S = \fint d\hat{s}^2 = \fint D^{-2}
\end{equation}
a definition that holds whether or not an underlying spacetime manifold is present.  When a manifold is present, this spectral action can be shown to be equal to the Einstein-Hilbert action, in the following manner. The non-commutative integral $\fint d\hat{s}^2 = \fint D^{-2}$ is given by the Wodzicki residue $Res_{W}D^{-2}$, which in turn is proportional to the volume integral of the second coefficient in the heat kernel expansion of $D^{2}$. The Lichnerowicz formula relates the square of the Dirac operator to the scalar curvature, thus enabling the remarkable result \cite{Landi1999}
\begin{equation}
\fint d\hat{s}^2 = - \frac{1}{48\pi^2} \int_{M} d^4x\; \sqrt{g} \ R
\end{equation}
In connection with  the standard model of particle physics coupling to gravity, the spectral action of the gravity sector can be written as a simple function of the square of the Dirac operator, using a cut-off function $\chi(u)$ which vanishes for large $u$ (\cite{Landi1999} and references therein):
\begin{equation}
S_G[D] =  \kappa Tr [\chi (L_P^2 D^2)]
\label{spect}
\end{equation} 
The constant $\kappa$ is chosen so as to get the correct dimensions of action, and the correct numerical coefficient. 

At curvature scales smaller than Planck curvature, this action can be related to the Einstein-Hilbert action using the following well-known heat kernel expansion:
\begin{equation}
S_G[D] = L_P^{-4} f_0 \; \kappa \int_M d^4x\ \sqrt{g} + L_p^{-2} f_2 \kappa \int_M d^4x \sqrt{g} R + ...
\label{sgd}
\end{equation}
Here, $f_0$ and $f_2$ are known functions of $\chi$ and the further terms which are of higher order in $L_p^{2}$ are ignored for the present. Also, we will not consider the cosmological constant term for the purpose of the present discussion. The development of a full quantum theory of gravity must take into account all higher order corrections. The present program is a truncated approximation to such a future theory.

Let us compare and contrast the above definition of spectral action with how a trace action is defined in Adler's theory of trace dynamics. In trace dynamics it is the Lagrangian [not the action] which is made of trace of a polynomial. Thus, the way things stand, we cannot use the spectral action directly in trace dynamics to bring in gravity into matrix dynamics. We need to think of the spectral action as a Lagrangian, and we then need to integrate that Lagrangian over time, to arrive at something analogous to the action in trace dynamics. We can convert the spectral action into a quantity with dimensions of a Lagrangian, simply by multiplying it by $c/L_p$ (equivalently, dividing by Planck time). But which time parameter to integrate the Lagrangian over? The space-time coordinates have already been assumed to be non-commuting operators, (especially in the definition of the atom of space-time-matter, as below, the case that we are interested in). So it seems as if we have a Lagrangian, but we do not have a time parameter over which to integrate the Lagrangian, so as to make an action.
Fortunately, non-commutative geometry itself comes with a ready-made answer! The required time parameter is the Connes time $\tau$, which we discussed in earlier work. In NCG, according to the 
Tomita-Takesaki theorem, there is a one-parameter group of inner automorphisms of the algebra ${\cal A}$ of the non-commuting coordinates - this serves as a `god-given' (as Connes puts it) time parameter with respect to which non-commutative spaces evolve \cite{Connes2000}. This Connes time $\tau$ has no analog in the commutative case, and we employ it here to describe evolution in trace dynamics. Thus we define the action for gravity, in trace dynamics, as
\begin{equation}
S_{GTD} =\kappa \frac{c}{L_P} \int d\tau \; Tr [\chi (L_P^2 D_B^2)]
\end{equation}
Note that $S_{GTD}$ has the correct dimensions, those of action. Also, we will henceforth denote the standard Dirac operator as $D_B$, instead of as $D$.

Next, we  derive the Lagrange equations for this trace action. For this we need to figure out what the configuration variables $q$ are. In the presence of a manifold, those variables would simply be the metric. But we no longer have that possibility here. We notice though that the operator $D_B$ is like momentum, since it has dimensions of inverse length. $D_B^2$ is like kinetic energy, so its trace is a good candidate Lagrangian. Therefore, we define a new self-adjoint  bosonic operator $q_B$, having the dimension of length, and we define a velocity $dq_B/d\tau$, which is defined to be related to the Dirac operator $D_B$ by the following new relation
\begin{equation}
D_B \equiv \frac{1}{Lc}\;  \frac{dq_B}{d\tau}
\end{equation}
where $L$ is a length scale whose significance will become clear shortly. The action for gravity in trace dynamics can now be written as
\begin{equation}
S_{GTD} =\kappa \frac{c}{L_P} \int d\tau \; Tr [\chi (L_P^2 \dot{q}^2/L^2 c^2)]
\end{equation}
where the time derivative in $\dot{q}$ now indicates derivative with respect to Connes time. For the present we will work with the function $\chi(u)=u$, leaving the consideration of convergence for future work.

We would now like to incorporate matter fermions into the theory. However we do not write the standard Dirac action for fermions, add up over all the fermions, and add that action to the gravity trace action. This is because at the Planck scale, where this theory operates, we do not make a distinction between a material particle described by a fermionic operator $q_F$, and its associated gravity $q_B$. Rather, we define an `atom of space-time-matter [STM]' [equivalently, an aikyon] by a Grassmann operator $q$ such that $q=q_B + q_F$. The natural split of $q$ into its bosonic and fermionic parts is equivalent to considering the aikyon as a combination of its matter part and its gravity part. The Hilbert space of the theory is populated by many STM atoms, each with its own operator $q_i$. The operator $q_F$ of an STM atom is used to define the `fermionic' Dirac operator $D_F$:
\begin{equation}
 D_B \equiv \frac{1}{Lc}\;  \frac{dq_F}{d\tau}
 \end{equation}
$D_B$ is defined such that in the commutative limit, it becomes the standard Dirac operator on a Riemannian manifold. $D_F$ is defined such that it gives rise to the classical action for a relativistic point particle, as we will see below.
An STM atom is assumed to be described by the following  action principle in this generalised trace dynamics including gravity:
 \begin{equation}
 \frac{S}{C_0}  =  \frac{1}{2} \int \frac{d\tau}{\tau_{Pl}} \; Tr \bigg[\frac{L_P^2}{L^2c^2}\; \left(\dot{q}_B +\beta_1 \frac{L_P^2}{L^2}\dot{q}_F\right)\; \left(\dot{q}_B +\beta_2 \frac{L_P^2}{L^2} \dot{q}_F\right) \bigg]
\end{equation}
where $\beta_1$ and $\beta_2$ are  constant self-adjoint fermionic matrices. These matrices make the Lagrangian bosonic.
The only two fundamental constants are Planck length and Planck time - these scale the length scale $L$ of the STM atom, and the Connes time, respectively. $C_0$ is a constant with dimensions of action, which will be identified with Planck's constant in the emergent theory. The Lagrangian and action are not restricted to be self-adjoint.

The canonical momenta obtained from this Lagrangian are constant and are given by
\begin{align}
    p_B = \frac{\delta \textbf{L}}{\delta \dot{q}_B} &= \frac{a}{2}\bigg[2\dot{q}_B + \frac{L_P^2}{L^2}(\beta_1 +\beta_2)\dot{q}_F \bigg] = c_1\\ 
    p_F = \frac{\delta \textbf{L}}{\delta \dot{q}_F} &= \frac{a}{2} \frac{L_P^2}{L^2}\bigg[\dot{q}_B (\beta_1 +\beta_2)+ \frac{L_P^2}{L^2}\beta_1 \dot{q}_F \beta_2 +  \frac{L_P^2}{L^2}\beta_2 \dot{q}_F \beta_1 \bigg]=c_2
\end{align}
where $a\equiv L_P^2/L^2c^2$.
These equations can be integrated to obtain the following solutions:
\begin{align}
     \dot{q}_B &= \frac{1}{2}\bigg[c_1 -(\beta_1+\beta_2)(\beta_1-\beta_2)^{-1}\big[2c_2 -c_1(\beta_1+\beta_2) \big](\beta_2-\beta_1)^{-1} \bigg] \label{qb} \\
     \dot{q}_F &= (\beta_1-\beta_2)^{-1}\big[2c_2-c_1(\beta_1+\beta_2)\big](\beta_2-\beta_1)^{-1} \label{qf}
\end{align}
This means that the velocities $\dot{q}_B$ and $\dot{q}_F$ are constant,  and $q_B$ and $q_F$ evolve linearly in Connes time. 
The trace Hamiltonian is given by
\begin{equation}
    \textbf{H} = \text{Tr} \frac{2}{a} \bigg[(p_B\beta_1-p_F)(\beta_2-\beta_1)^{-1}(p_B\beta_2-p_F)(\beta_1-\beta_2)^{-1}
    \bigg]
\end{equation}
The Adler-Millard charge is given by
\begin{align}
     (2/a) \; \tilde{C} &= [q_B, 2\dot{q}_B +(\beta_1+\beta_2)\dot{q}_F] -\{q_F, \dot{q}_B(\beta_1 +\beta_2)+\beta_1 \dot{q}_F \beta_2+ \beta_2 \dot{q}_F \beta_1 \} \nonumber\\
      &= [q_B,2\dot{q}_B]+[q_B,(\beta_1+\beta_2)\dot{q}_F]-\{q_F,\dot{q}_B(\beta_1+\beta_2)\} \\ \nonumber &\quad \: -\{q_F,\beta_1\dot{q}_F\beta_2+\beta_2\dot{q}_F\beta_1 \} 
\end{align}
[In Eqns. (30-33) we have suppressed the factor $L_P^2/L^2$ so as to keep the expression from being complicated; it is understood that every $\beta$ in these equations comes multiplied with this factor.]
The equation for the bosonic momentum $p_B$ can be written as a modified Dirac equation with a complex eigenvalue:
\begin{equation}
\left[D_B + \frac{L_P^2}{L^2}\frac{\beta_1+\beta_2}{2}D_F\right] \psi =  \frac{1}{L} \bigg(1+ i \frac{L_P^2}{L^2}\bigg)\psi
\label{mod}
\end{equation}
Since $D_B$ is self-adjoint, the imaginary part to the eigenvalue comes only from $D_F$, and the relative magnitudes of the real and imaginary part are dictated by the structure of the operator on the left. The eigenvector depends on both $q_B$ and $q_F$. This equation plays a crucial role in the subsequent discussion below. 
We note that $D_F$ will also contribute a term to the real part of the eigenvalue; let us denote this term by $\frac{L_P^2}{L^2}\theta$. In the limit $L\gg L_P$, this term is negligible. It turns out this will be the quantum limit: the imaginary part of the eigenvalue is also ignorable, and one effectively has a self-adjoint trace Hamiltonian. In the limit $L\ll L$ this term will be significant - this happens to be the classical limit: there is also a non-negligible imaginary `fast' component at the Planck scale, which gives rise to a significant anti-self-adjoint part in the Hamiltonian. It is not clear to us at present as to what role the $\theta$ term is playing in the classical limit. It appears to modify classical general relativity, but does not affect our subsequent calculation of the black hole entropy.

We have now described our trace dynamics model including gravity. If there are $N$ aikyons in the system, the above action is written for each aikyon ($L$ can be different in magnitude for different aikyons), and the total action is the sum of the individual actions. We call this theory spontaneous quantum gravity.

The next step is to carry out the statistical thermodynamics for a large number of aikyons, and to understand the emergent quantum gravity theory, as well as the emergence of classical space-time geometry after spontaneous localisation. 
Consider first a collection of STM atoms each of which has the property $L\gg L_P$. Then the imaginary component of the eigenvalue in the modified Dirac equation becomes negligible. As a result the trace-Hamiltonian is self-adjoint, and the Adler-Millard charge is anti-self adjoint. This is also equivalent to justifiable neglect of the "fast" imaginary component of the dynamical variables $x_r$ at the Planck scale. Hence the conditions necessary for arriving at the equilibrium statistical thermodynamics by coarse-graining over large times are satisfied.

This sets the stage for the emergence of the averaged quantum gravitational dynamics at statistical equilibrium. A Ward identity, which is the equivalent of the equipartition theorem, is derived. As in trace dynamics, the anti-self adjoint part of the conserved Adler-Millard charge is equipartitioned over all the degrees of freedom, and the equipartitioned value per degree of freedom is identified with Planck's constant $\hbar$. (The constant $C_0$ is now identified with $\hbar$.) At equilibrium, the standard quantum commutation relations of (an equivalent of) quantum general relativity emerge, for the canonical ensemble averages of the various degrees of freedom:
\begin{equation}
[q_B, p_B] = i \hbar; \qquad \{q_{FS}, p^f_{FAS}\} = i\hbar; \qquad \{q_{FAS}, p^f_{FS}\} = i\hbar
\end{equation}
The subscript $S/AS$ denote self-adjoint and anti-self-adjoint parts of the dynamical variables. The superscript $f$ denotes
the fermionic part of the momentum $p_F$, being the part which depends on $q_F$ but not on $q_B$: i.e. $p_F^{f} = \beta_1 \dot{q}_F\beta_2 +\beta_2 \dot{q}_F \beta_1 $.
All the other commutators and anti-commutators amongst the canonical degrees of freedom vanish at thermodynamic equilibrium. The above set of commutation relations hold for every STM atom. We note that we describe quantum general relativity in terms of these $q$ operators, and not in terms of the metric and its conjugate momenta, which are emergent concepts of Levels II and III. There is likely a possible connection between this description of quantum general relativity, and loop quantum gravity, which remains to be explored.

The mass $m$ of the aikyon is {\it defined} by $m\equiv\hbar/Lc$; and as a consequence $L$ is hence interpreted to be its Compton wavelength. Newton's gravitational constant $G$ is defined by $G\equiv L_p^2 c^3/\hbar$, and Planck mass $m_P$ by $m_P=\hbar/L_P c$. Mass and spin are both emergent concepts of Level I; at Level 0 the aikyon only has an associated length $L$ - this length is a property of both the gravity aspect and the matter aspect of the STM atom.

As a consequence of Hamilton's equations for the matrix dynamics at Level 0, and as a consequence of the Ward identity mentioned above, the canonical ensemble averages of the canonical variables obey the Heisenberg equations of motion of quantum theory, these being determined by ${ H}_S$, the canonical average of the self-adjoint part of the Hamiltonian:
\begin{equation} 
i\hbar \frac{\partial q_B}{\partial \tau} = [q_B, { H}_S]; \qquad i\hbar \frac{\partial p_B}{\partial \tau} =  [p_B, { H}_S]; \qquad i\hbar \frac{\partial q_F}{\partial \tau} = [q_F,{ H}_S]; \qquad i\hbar \frac{\partial p^f_F}{\partial \tau} =  [p^f_F, { H}_S]
\end{equation}
In analogy with quantum field theory, one can transform from the above Heisenberg picture, and write a Schr\"{o}dinger equation for the wave-function $\Psi(\tau)$ of the full system:
\begin{equation}
i\hbar \; \frac{\partial \Psi}{\partial \tau} = { H_{Stot}} \Psi (\tau)
\end{equation}
where ${ H_{Stot}}$ is the sum of the self-adjoint parts of the Hamiltonians of the individual STM atoms. Since the Hamiltonian is self-adjoint the norm of the state vector is preserved during evolution.  This equation is the analog of the Wheeler-DeWitt equation in our theory, the equation being valid at thermodynamic equilibrium at Level I. This equation can possibly resolve the problem of time in quantum general relativity, because to our understanding it does not seem necessary that the physical state must be annihilated by ${ H_{Stot}}$. We have not arrived at this theory by quantising classical general relativity; rather the classical theory will emerge from here after spontaneous localisation, as we now describe.

It is known that the above emergence of quantum dynamics arises at equilibrium in the approximation that the Adler-Millard conserved charge is anti-self-adjoint, and its sef-adjoint part can be neglected. In this  approximation, the Hamiltonian is self-adjoint. Another way of saying this is that quantum dynamics arises when statistical fluctuations around equilibrium (which are governed by the self-adjoint part of $\tilde{C}$) can be neglected. These fluctuations arise when the "fast" component due to the imaginary eigenvalue in the modified Dirac equation becomes significant. This happens if $L\ll L_{Pl}$. For a single aikyon whose mass is much less than Planck mass, this would be impossible. Consider however a very large collection of aikyons which are entangled with each other [Level I description]. The effective Compton wavelength $L_{eff}$, as it would appear in the effective modified Dirac equation,  is then given by
\begin{equation} 
\frac{1}{L_{eff}} = \sum_i \frac{1}{L_i}
\end{equation}
Clearly, if a very large number of aikyons get entangled, their total mass can exceed Planck mass significantly; and the effective Compton wavelength becomes much smaller than Planck length. This is indicative of emergent classical behaviour, as follows. The fast varying imaginary component in the modified Dirac equation, on the Planck scale, is represented as imaginary stochastic corrections to the equilibrium quantum dynamics.

When the thermodynamical fluctuations are important, one must represent them by adding a stochastic anti-self-adjoint operator function to the total self-adjoint Hamiltonian (note that one cannot simply add the anti-self-adjoint part of the Hamiltonian to the above Schr\"{o}dinger equation, because that equation is defined for canonically averaged quantities; the only way to bring in fluctuations about equilibrium is to represent them by stochastic functions). This way of motivating spontaneous collapse is just as in trace dynamics (see Chapter 6 of \cite{Adler:04}), except that we are not restricted to the non-relativistic case, and evolution is with respect to Connes time $\tau$. Also, we do not have a classical space-time background yet; this will emerge now, as a consequence of spontaneous localisation [see also our earlier related paper `{\it Space-time from collapse of the wave function}' \cite{Singh:2019}].

Thus we can represent the inclusion of the anti-self-adjoint fluctuations in the above Schr\"{o}dinger equation by a stochastic function ${\cal H}(\tau)$ as:
\begin{equation}
i\hbar \; \frac{\partial \Psi}{\partial \tau} = [{ H_{Stot}} + {\cal H}(\tau)] \Psi (\tau)
\end{equation}
In general, this equation will not preserve norm of the state vector during evolution. However, as we noted above, every STM atom is in free particle motion. Hence it is  reasonable to demand that the state vector should preserve norm during evolution, even after the stochastic fluctuations have been added. Then, exactly as in collapse models and in trace dynamics, a new state vector is defined, by dividing $\Psi$ by its norm, so that the new state vector preserves norm. Then it follows that the new norm preserving state vector obeys an equation which gives rise to spontaneous localisation, just as in trace dynamics and collapse models (see Chapter 6 of \cite{Adler:04}). We should also mention that the gravitational origin of the anti-self-adjoint fluctuations presented here ($D_F$ is likely of gravitational origin, and relates to the anti-symmetric part of an asymmetric metric) agrees with Adler's proposal that the stochastic noise in collapse models is seeded by an imaginary component of the metric \cite{Adler2014, Adler:2018}.

It turns out to be instructive to work in the momentum basis where the state vector is labelled by the eigenvalues of the momenta $p_B$ and $p_F$. Since the Hamiltonian depends only on the momenta, the anti-self adjoint fluctuation is determined by the anti-self adjoint part of $p_F$. Hence it is reasonable to assume that spontaneous localisation takes place onto one or the other eigenvalue of $p^f_F$. No localisation takes place in $p_B$ - this helps understand the long range nature of gravity (which results from $q_B$ and the bosonic Dirac operator $D_B$). We assume that the localisation of $p^f_F$ is accompanied by the localisation of $q_F$, and hence that an emergent classical space-time 
is defined using the eigenvalues of $q_F$ as reference points.  Space-time emerges only as a consequence of the spontaneous localisation of matter fermions. Thus we are proposing that the eigenvalues of $q_F$ serve to define the space-time manifold. 

We need to ask how a space-time manifold emerges after spontaneous localisation of fermions. Localised fermions serve as physical markers of space-time points, in the spirit of the Einstein hole argument. The recovery of the standard (commutative) Riemannian manifold is achieved because spontaneous localisation undoes the process
[${\rm Space-time\ points} \rightarrow {\rm Operators}$] achieved by going from a commutative algebra of coordinates to a non-commutative algebra. Thus, to begin with, there is a Riemannian geometry on a space-time manifold [assumed to be four-dimensional]; it is mapped to a commutative algebra, including a (diffeomorphism invariant) algebra of coordinates. When this algebra is made non-commutative, geometric concepts such as distance, metric and curvature can still be preserved, by employing the Dirac operator $D_B$. In our theory with STM atoms, each atom is by itself a non-commutative geometry, complete with these concepts. This NCG has been arrived at by raising each space-time point to operator/matrix status. What spontaneous  collapse does is to dynamically reverse this process, and restore space-time operators back to points. If sufficiently many STM atoms undergo localisation, then the manifold, metric and curvature concepts are recovered. The classical space-time manifold acts as a boundary condition which has to be fulfilled by the matrix dynamics. The space-time coordinates and metric which were present before the lift to the non-commutative case is restored.

 As in objective collapse models, the rate of spontaneous  localisation becomes significant only for objects which consist of a large number of matter fermions - hence the emergence of a classical space-time is possible only when a sufficiently macroscopic object comprising many STM atoms undergoes spontaneous localisation. The rate of localisation $T$ is in fact given by $T=
 \hbar^2/GM^3 c$ where $M$ is the total mass of the macroscopic entangled system.
  We now give a quantitative estimate as to what qualifies as being sufficiently macroscopic. 

To arrive at these estimates, we recall the following two earlier equations, the action principle for the aikyon itself, and the eigenvalue equation for the full Dirac operator $D$:
\begin{equation}
\frac{L_P}{c} \frac{S}{\hbar}  =  \frac{a}{2} \int d\tau \; Tr \bigg[ \dot{q}_B^2 +\dot{q}_B\frac{L_P^2}{L^2}\beta_2\dot{q}_F +\frac{L_P^2}{L^2}\beta_1 \dot{q}_F \dot{q}_B +\frac{L_P^4}{L^4}\beta_1 \dot{q}_F \beta_2 \dot{q}_F \bigg]
\label{newacnr}
\end{equation}
\begin{equation}
[D_B + \frac{L_P^2}{L^2}\frac{\beta_1+\beta_2}{2}D_F] \psi = \lambda \psi \equiv (\lambda_R + i \lambda_I)\psi \equiv \bigg(\frac{1}{L} + i \frac{1}{L_I}\bigg)\psi
\end{equation}
In the second equation, since $D$ is bosonic, we have assumed that the eigenvalues $\lambda$ are complex numbers, and separated each eigenvalue into its real and imaginary part.  Recall that $L_I=L^3/L_P^2$. There is one such pair of equations for each aikyon, and the total action of all aikyons will be the sum of their individual actions, with the individual action as given above.

When an aikyon undergoes spontaneous localisation, $p^f_F$ localises to a specific eigenvalue. Since $D_F$ is also made from $\dot{q}_F$, just as $p^f_F$ is, we assume that $D_F$ also localises to a specific eigenvalue, whose imaginary part is the $L_I$ introduced above. Correspondingly, the $D_B$ associated with this STM atom acquires a real eigenvalue, which we identify with the $\lambda_R\equiv 1/L$ above (setting aside for the moment the otherwise plausible situation that in general $p_F$ will also contribute to $\lambda_R$). 

The spontaneous localisation of each aikyon to a specific eigenvalue reduces the first term of the trace Lagrangian to:
\begin{equation}
Tr [\dot{q}_B^2] \rightarrow \lambda_R^2
\end{equation}
If sufficiently many aikyons undergo spontaneous localisation to occupy the various eigenvalues $\lambda^i_R$  of the Dirac operator $D_B$, then we can conclude, from our knowledge of the spectral action in non-commutative geometry \cite{Landi1999}, that their net contribution to the trace is:
\begin{equation} 
\frac{\hbar a}{2} \;  Tr[\dot{q}_{B}^2] =\frac{\hbar }{2}  Tr [L_p^2 D_B^2] = \frac{\hbar}{2} L_p^2 \sum  (\lambda_R^i)^2 = \frac{\hbar}{2 L_p^2} \int d^4x \; \sqrt{g} \; R
\end{equation}
Thus we conclude that the Einstein-Hilbert action emerges after spontaneous localisation of the matter fermions. In that sense, gravitation is indeed an emergent phenomenon. Also, the eigenvalues of the Dirac operator $D_B$ have been proposed as dynamical observables for general relativity \cite{Rovelli}, which in our opinion is a result of great significance.
This study also demonstrates how to relate the eigenvalues of $D_B$ to the classical metric. In this sense the matrix $q_B$ captures the information of the metric field.

Let us now examine how the matter part of the general relativity action arises from the trace Lagrangian (its second and third terms) arises after spontaneous localisation. These terms are given as
\begin{equation}
\frac{L_p^2}{L^2}\frac{a\hbar }{2}  Tr \big[ \dot{q}_B\beta_2\dot{q}_F +\beta_1 \dot{q}_F \dot{q}_B \big] = \hbar Tr [L_p^2 \times \frac{L_P^2}{L^2}\frac{\beta_1+\beta_2}{2}D_F D_B]
\end{equation}
Spontaneous localisation sends this term to $L_p^2 \times {1}/{L_I} \times {1}/{L}$, where $L_I=L^3/L_P^2$. There will be one such term for each STM atom, and analogous to the case of $Tr D_B^2$ we anticipate that the trace over all STM atoms gives rise to the `source term'
\begin{equation}
\hbar\int \sqrt{g} \; d^4x \;\sum_i [ L_p^{-2} \times {1}/{L^i_I} \times {1}/{L^i}]
\end{equation}
Consider the term for one aikyon. We make the plausible assumption that spontaneous localisation localises the STM atom to a size $L_I$. This is analogous to the resolution length scale (conventionally denoted as $r_c$ in collapse models).  We know that $L_p^2 L_I = L^3$. We recall that  $L$ is the Compton wavelength $\hbar/ mc$ of the aikyon. Moreover, we propose that the classical approximation consists of replacing the inverse of the spatial volume of the localised particle - $1/L^3$, by the spatial delta function $\delta^3({\bf x} - {\bf x_0})$ so that the contribution to the matter source action becomes
\begin{equation}
\hbar \int \sqrt{g} \; d^4x \; [ L_p^{-2} \times {1}/{L_I} \times {1}/{L}] = mc \int ds
\end{equation}
which of course is the action for a relativistic point particle.

Putting everything together, we conclude that upon spontaneous localisation, the fundamental trace based action for a collection of aikyons becomes
\begin{equation}
S = \int d^4x\; \sqrt{g} \; \bigg [\frac{c^3}{2G}R + c\; \sum_i m_i \delta^3({\bf x} - {\bf x_0})\bigg]
\end{equation}
In this way, we recover general relativity at Level III, as a result of spontaneous localisation of quantum general relativity at Level I. We should not think of the gravitational field of the STM atom as being disjoint from its related fermionic source: they both come from the same eigenvalue $\lambda$, being respectively the real and imaginary parts of this eigenvalue.

We can now explain why each of the point mass localisations represents a Schwarzschild black hole. The localisation takes place to a size $L_I=L^3/L_P^2$ and since the mass of the entangled atoms is much higher than Planck mass, and since $L$ is its effective Compton wavelength, $L_I$ is much smaller than Planck length. Thus we have a point mass like solution confined to below Planck length, which we have plausibly approximated by a delta function. The associated Schwarzschild radius $L_P^2/L$ is much greater than Planck length, implying that localisation happens much inside the Schwarzschild radius. The gravitational field of such a matter source is described by the emergent Einstein equations written above and is hence a Schwarzschild black hole. We note that spontaneous localisation is a process different in nature from classical gravitational collapse. Since the mass of the macroscopic object is Planck mass or higher, repeated spontaneous localisation to the same location keeps taking place at an extremely rapid rate, keeping the object as a classical black hole. In the next section we will calculate the entropy of one such black hole. On the other hand, those entangled particles whose total mass is less than Planck mass, remain quantum after spontaneous localisation [i.e. do not form a black hole] because the Compton wavelength exceeds Schwarzschild radius. Thus there is a transition from classical black hole phase to quantum phase, when the net entangled mass becomes larger than Planck mass. Since there are no non-gravitational forces in our theory, spontaneous localisation of massive objects necessarily forms black holes. As and when these other interactions are included, spontaneous collapse would give rise to ordinary [non-black-hole] macroscopic objects. Interestingly, Planck length becomes the minimum observable length, because when the Compton wavelength $L$ is smaller than Planck length, the associated Schwarzschild radius exceeds Planck length, and is the observable size. At Planck mass, the Schwarzschild radius and Compton length are both equal to Planck length  this being the minimum observable length.

\section{Including Yang-Mills fields}
Our theory thus far includes gravity and Dirac fermions, both described by one common term in the action principle for the aikyon. It is incomplete because it does not include Yang-Mills gauge fields. Fortunately, it turns out that it is not difficult to include gauge fields in the action for the aikyon. We recall that in quantum mechanics the gauge potential is included as a correction to the momentum, i.e. as a correction to the Dirac operator [why this should be so becomes clearer from our recently proposed definition of spin angular momentum in this matrix dynamics: the matrix dynamics combines linear angular momentum and spin angular momentum into real and imaginary parts of a complex momentum, respectively]. Since the Dirac operator is represented as the gravitational aspect $\dot{q}_B$ of the aikyon, we represent Yang-Mills gauge field as $q_B$, and the associated fermionic current as $q_F$. 
Thus we defined the modified Dirac operators $D_{Bnewi}$ and $D_{Fnewi}$ by:
\begin{equation}
{\dot{\widetilde{Q}}_B} = \frac{1}{L} (i\alpha q_B + L \dot{q}_B); \qquad  {\dot{\widetilde{Q}}_F} = \frac{1}{L} (i\alpha q_F + L \dot{q}_F);
\end{equation}
\begin{equation}
D_{Bnewi} = \dfrac{1}{L} \dot{\widetilde{Q}}_{B} \qquad and \qquad D_{Fnewi} = \frac{L_P^2}{L^2}\frac{\beta_1+\beta_2}{2Lc} \dot{\widetilde{Q}}_{F}
\label{eq:ddirac}
\end{equation}
Here, $\alpha$ is the gauge coupling constant. The new Lagrangian is  given by \cite{mps}
\begin{equation}
\mathcal{L} = Tr \biggl[\biggr. \dfrac{L_{p}^{2}}{L^{2}} \biggl(\biggr. \dot{\widetilde{Q}}_{B} + \dfrac{L_{p}^{2}}{L^{2}} \beta_{1} \dot{\widetilde{Q}}_{F} \biggl.\biggr) \biggl(\biggr. \dot{\widetilde{Q}}_{B} + \dfrac{L_{p}^{2}}{L^{2}} \beta_{2} \dot{\widetilde{Q}}_{F} \biggl.\biggr) \biggl.\biggr]
\end{equation}
It is highly significant that this Lagrangian has the same structure as for pure gravity, except that the dynamical variable has now become complex. As if to suggest that after including non-gravitational interactions it is still possible to obtain a geometric picture. 
The result from geometry relating $Tr[D_B]^2$ to the Einstein-Hilbert action can be generalised to include Yang-Mills fields. It is possible that these complex dynamical variables represent a complex metric with an imaginary anti-symmetric part, and that the latter relates to torsion and in turn to Yang-Mills gauge fields and spin. This important possibility is currently under investigation. The space-time symmetry group would then be a non-commutative and complex generalisation of the Poincare group; with the Lorentz group part relating to gravity, and the translations relating to spin / torsion / gauge-fields. The inclusion of Yang-Mills fields also facilitates an understanding of quantum spin,  and of the spin-statistics connection \cite{Singhspin}. 

The analysis of the equations of motion, the statistical thermodynamics, spontaneous localisation and the classical limit proceeds along the same lines as for the pure gravity case. Significantly, we employed this analysis to show that the Kerr-Newman black has the same gyromagnetic ratio as the electron [both being twice the classical value], an intriguing fact for which there was no convincing explanation so far \cite{mps}


\section{Physical applications and predictions of Spontaneous Quantum Gravity}
Since our theory of quantum gravity is intimately connected to fermions and Yang-Mills fields, and to the low energy universe, it makes several falsifiable predictions which can be used to confirm or rule out the theory. These are:

1. Spontaneous localisation (the GRW theory) is a prediction of this theory, and the GRW theory is being tested in labs currently. If the GRW theory is ruled out by experiments, our  proposal will be ruled out too.

2. We have predicted the novel phenomena of quantum interference in time, and spontaneous collapse in time. This is discussed in \cite{Singh:2019}  and is falsifiable.

3. We have given a derivation of the Bekenstein-Hawking entropy of the Schwarzschild black hole, from the microstates of the aikyons that make up the black hole \cite{maithresh2019b}.

4. The theory predicts the Karolyhazy length as a minimum length, as a consequence of the relation between $L$ and $L_I$. This is testable and falsifiable, as discussed in \cite{SinghQGV2019}.

5. This theory predicts that dark energy is a quantum gravitational phenomenon, as discussed in \cite{Singh:DE}.

6. We explain why the Kerr-Newman black hole has the same gyromagnetic ratio as the electron, both being twice the classical 
value \cite{mps}.

We are currently investigating if our theory can accommodate the standard model of particle physics, or at least explain some of its features.

\section{A comparison with other quantum gravity approaches}
Our theory has been built on the following foundational principle: there ought to exist a reformulation of quantum theory which does not refer to classical time. This naturally leads us to the aforesaid deterministic matrix dynamics on the Planck scale, building on the theories of trace dynamics and non-commutative geometry. We recover quantum field theory, and classical general relativity, as  low energy approximations below the Planck scale. The absence of superpositions of space-time geometries in the classical limit is explained dynamically, because the original Hamiltonian of matrix dynamics is not self-adjoint. Quantum theory is recovered when the anti-self-adjoint part is negligible. Spontaneous localisation and classical limit is recovered when the anti-self-adjoint part is significant. 

These features make our approach to quantum gravity quite different from the other existing approaches. Many of the existing approaches accept the validity of quantum field theory at the Planck scale. This then necessitates that the recovery of the classical limit is through one or the other interpretations of quantum mechanics, such as the many-worlds interpretation, consistent histories, Bohmian mechanics, or one of the other interpretations. Thus, is it dynamical spontaneous localisation, or a quantum interpretation, which leads to classicality? This is the most striking difference: whether spontaneous localisation is right or not can be settled by experimental tests.

Leaving aside for the moment the issue of classicality, there is perhaps interesting commonality with some of the other approaches to quantum gravity. The action principle for the aikyon, which unifies gravity, gauge fields and fermions, could bear a similarity to the classical closed string of string theory, and this is an aspect worth investigating further. The quantum gravity theory that emerges from our matrix dynamics bears resemblance to the Wheeler-deWitt equation as well as to loop quantum gravity, and these aspects are worth investigating further.

Our work resonates strongly with attempts to incorporate the standard model of particle physics in Connes' non-commutative geometry \cite{Schucker}.
In fact our approach is strongly inspired by their description of geometry [both Riemannian and non-commutative] in terms of the Dirac operator. Without this finding of theirs, our theory would not stand. We used this result to bring gravity within the fold of trace dynamics. It now remains to be seen if the standard model can be explained from our Planck scale matrix dynamics.

There have been other interesting `quantum-first' approaches to gravity, for example the work of Giddings \cite{Giddings1, g2} and Carroll and collaborators \cite{Carroll1, c2}. The idea here is that instead of quantising an already given classical theory of gravity, one looks to add fundamental structure to quantum mechanics, which would enable the inclusion of gravity [in a quantum gravity sense], and from which classical space-time geometry will be emergent, possibly as a consequence of entanglement. There are important commonalties between these approaches, and ours. The common goal is that something should be done to quantum mechanics so as to include gravity in it, and also to make key use of entanglement. What we `do' to quantum theory is to remove classical space-time from it. However, there are important differences too, from these approaches. These approaches would like to retain the concepts of unitarity and locality/separability. The matrix dynamics we construct is non-unitary, with unitary quantum field recovered as a low energy approximation, in the limit of sub-critical entanglement. Also, the matrix dynamics is separable in the sense that the different STM atoms are enumerable, but the dynamics is not local, in the sense that space-time and matter are not distinct from each other. Classical space-time, locality, and material separability are recovered in a low energy approximation, as a consequence of super-critical entanglement. The presence of an anti-self-adjoint part in the matrix Hamiltonian assists the entangled STM atoms to undergo dynamical spontaneous localisation, giving rise to emergent locality, separability, and classicality. This same non-unitary aspects permits the Karolyhazy relation to arise, and hence also the quantum gravitational dark energy. To our understanding, in the other quantum-first gravity approaches, the status of the cosmological constant and vacuum energy  is not changed.

In spirit and philosophy, our work is closest to the theory of emergent quantum mechanics being developed by Torrome \cite{Torrome}. We are currently exploring the points of agreement and differences in our approaches.

\section{Concluding Remarks}     
      
Quantum gravity, in its most fundamental sense at Level 0, is a deterministic matrix dynamics of STM atoms \cite{Singhdice}. These interact with each other via entanglement. Hence entanglement is more fundamental than quantum theory, and it is first and foremost a property of STM atoms evolving in Hilbert space. This also makes it very clear why quantum entanglement is oblivious to space and time (quantum non-locality) - because entanglement originates from Level 0 and Level I, where there is no space-time. Quantum dynamics should strictly be described at Level 0 or Level I. Describing it at level II is an approximation; this can sometimes lead to puzzles - for instance the EPR paradox arises when we try to describe quantum non-locality at Level II. There is no EPR paradox at Levels 0 and I, because there is no space-time there, so there is no question of a space-like separation. Space-time is emergent from Hilbert space, after spontaneous localisation takes place. 

Any quantum theory of gravity must also explain why superpositions of space-time geometries are absent in the classical world. Moreover, the absence of position superpositions of macroscopic bodies is a pre-requisite for the existence of classical space-time geometry. In this way of thinking it becomes apparent that the solution of the quantum measurement problem must come from a quantum theory of gravity. Since our quantum gravity predicts spontaneous localisation of fermions, we see that the process responsible for the emergence of space-time is the same as the one that solves the measurement problem.

In order to have a relativistic theory of spontaneous collapse, it is necessary to treat time at the same footing as space. This requires that just like the position operator, time should also be treated as an operator - then there is spontaneous collapse of time as well. The loss of coordinate time as a parameter is compensated by the appearance of Connes time as the  new time parameter.

Many researchers have made the case that gravity is not a fundamental force, but an emergent thermodynamic phenomenon (Sakharov, Jacobson, Padmanabhan, Verlinde, amongst others). There are underlying atoms of space-time. Adler has made the case that quantum theory is an emergent phenomenon. We agree with both these cases and we have 
made the case that quantum gravity itself is an emergent phenomenon, coming from the matrix dynamics of STM atoms. Space-time and its geometry, as well as the phenomenon of gravitation, emerge after the spontaneous localisation of the fermionic part of STM atoms. 
The thermodynamic properties of black holes testify for the emergent nature of gravity; while the random nature of outcomes in a quantum measurement testifies for the thermodynamic nature of quantum theory. In fact, the same process, viz. spontaneous localisation, explains the origin of black hole entropy, and also the collapse of the wave function in a quantum measurement.

Our underlying matrix dynamics is a deterministic and time-reversible theory; it is even linear!
The apparent irreversible nature of wave function collapse, as well as of black hole evaporation, arises only because we are examining a coarse-grained version of the matrix dynamics. It would have been hard to anticipate that the sought for quantum theory of gravity will turn out to the statistical thermodynamics corresponding to the microscopic dynamics of STM atoms. In hindsight though, it seems obvious that it should be so, because both gravity and quantum theory exhibit strong thermodynamic features. Quantum gravity is to the matrix dynamics of STM atoms same as the thermodynamic properties of a gas are to the mechanical motion of its constituent microscopic molecules.

Instead of developing this story in the top-down fashion as we have done in this article, one can now also describe it in a bottom-up fashion, by starting at the most basic Level 0. We start with the action principle for STM atoms and work out their Lagrangian dynamics. The statistical mechanics of these atoms gives rise to quantum gravity, and by spontaneous collapse, to classical general relativity with matter sources. Quantum field theory is arrived at 
by borrowing quantum matter from quantum gravity, and classical space-time from Level III.

\bigskip

I would like to thank Angelo Bassi, Hendrik Ulbricht and the other organisers of the Trieste TEQ Workshop `Redefining the foundations of physics in the quantum technology era' for bringing together researchers working on collapse models. I would also like to thank Universita Degli di Trieste for its kind hospitality, and for providing excellent facilities and atmosphere. The blue of the Adriatic Sea can do no less than to deeply inspire physicists to search for a pebble or two prettier than the ordinary. It is also my pleasure to thank Palemkota Maithresh, Meghraj M S and Abhishek Pandey for the collaboration that led to this new quantum theory of gravity.

\setstretch{1.25}

\bigskip
\bigskip

\centerline{\bf REFERENCES}

\bibliography{biblioqmtstorsion}

\end{document}